\documentclass[aps,prb,twocolumn,superscriptaddress,showpacs,showkeys ]{revtex4-1}
\usepackage{axodraw4j}
\usepackage{pstricks}
\usepackage{graphicx}
\usepackage{hyperref}
\usepackage{amsmath,amssymb}
\usepackage{epstopdf}
\usepackage{subcaption}
\usepackage[text={7in,10in}]{geometry}

\graphicspath{{figs/}}

\begin{document}
	
\title{Interaction-disorder-driven characteristic momentum in graphene, \\
approach of multi-body distribution functions}

\author{M. N. Najafi}
\affiliation{Department of Physics, University of Mohaghegh Ardabili, P.O. Box 179, Ardabil, Iran}
\email{morteza.nattagh@gmail.com}

\date{\today}

\begin{abstract}
Multi-point probability measures along with the dielectric function of Dirac Fermions in mono-layer graphene containing particle-particle and white-noise (out-plane) disorder interactions on an equal footing in the Thomas-Fermi-Dirac approximation is investigated. By calculating the one-body carrier density probability measure of the graphene sheet, we show that the density fluctuation ($\zeta^{-1}$) is related to the disorder strength ($n_i$), the interaction parameter ($r_s$) and the average density ($\bar{n}$) via the relation $\zeta^{-1}\propto r_sn_i^2\bar{n}^{-1}$ for which $\bar{n}\rightarrow 0$ leads to strong density inhomogeneities, i.e. electron-hole puddles (EHPs), in agreement with the previous works. The general equation governing the two-body distribution probability is obtained and analyzed. We present the analytical solution for some limits which is used for calculating density-density response function. We show that the resulting function shows power-law behaviors in terms of $\zeta$ with fractional exponents which are reported. The disorder-averaged polarization operator is shown to be a decreasing function of momentum like ordinary 2D parabolic band systems. It is seen that a disorder-driven momentum $q_{\text{ch}}$ emerges in the system which controls the behaviors of the screened potential. We show that in small densities an instability occurs in which imaginary part of the dielectric function becomes negative and the screened potential changes sign. Corresponding to this instability, some oscillations in charge density along with a screening-anti-screening transition are observed. These effects become dominant in very low densities, strong disorders and strong interactions, the state in which EHPs appear. The total charge probability measure is another quantity which has been investigated in this paper. The resulting equation is analytically solved for large carrier densities, which admits the calculation of arbitrary-point correlation function.
\end{abstract}


\maketitle

\section{Introduction}
Since the discovery of graphene as a hybrid between metal and insulator, this material with relativistic energy spectrum of zero-gap Dirac fermions \cite{wallace1947band,novoselov2004electric} attracts attention due to its unique electronic properties and prospective applications in nanoelectronics \cite{beenakker2008colloquium,neto2009electronic,sarma2011electronic,kotov2012electron}. The understanding of the origin and the influence of disorder, as well as interactions in graphene seems to be essential in understanding of the experiments and also in designing graphene-based electronic devices. The effect of these quantities and the resulting screening are not as easygoing as ordinary simple metals \cite{sarma2011electronic}. There is a huge literature concerning the interplay of particle interactions and disorder in mono and multilayer graphene \cite{stauber2005disorder,ostrovsky2006electron,aleiner2006effect,fogler2008effect,pereira2008modeling,mucciolo2010disorder,sadhukhan2017disorder,neto2009electronic} which cause variety of phenomenons. Examples of the effect of disorder on the properties of graphene are its effect on: the compressibility \cite{asgari2008effect}, the electron-phonon interaction \cite{song2012disorder}, the Fano factor and conductivity \cite{Lewenkopf2008numerica,sarma2011electronic}, the magnetoresistance of bilayer graphene \cite{hatami2011conductance}, the graphene Hall bars \cite{petrovic2016quantum}, the optical properties of graphene quantum dots \cite{altintacs2017effects} and the effective electronic mass in bilayer graphene \cite{li2016effective}. Disorder-based spintronics in graphene \cite{rocha2010disorder}, and Levi-flight transport caused by anisotropically distributed on-site impurities \cite{gattenlohner2016levy} are other examples for which the disorder plays a dominant role. The disorder-mediated Kondo effect \cite{miranda2014disorder} and the formation of electron-hole puddles (EHPs) are examples for which the disorder and the inter-particle interactions play vital roles simultaneously.  EHPs are believed to be responsible for the observed minimum conductivity of graphene and was predicted theoretically by Hwang \textit{et al} \cite{hwang2007carrier} and Adam \textit{et al} \cite{adam2007self}  and was also confirmed in experiments in the vicinity of the Dirac point \cite{martin2008observation,rutter2007scattering,brar2007scanning,zhang2009origin,deshpande2009spatially,deshpande2009mapping,ishigami2007atomic,cho2008charge,berezovsky2010imaging,berezovsky2010imaging2}. They are characterized by the state in which some strong carrier density inhomogeneities with density fluctuations much larger than the average density (for low densities) emerge \cite{rossi2008ground}. In this case (for which the transport is governed by the complex network of small random puddles with semi-metal character \cite{rossi2008ground}) and the similar situations, a many body treatment is needed in which the effect of interaction and the disorder enter to the play with the same footing. Such an investigation can be found in \cite{herbut2008coulomb} in which it has been shown that for the gauge field randomness, related to the formation of ripples characterized by a disorder strength $\Delta$, displaces the IR fixed point to $\alpha^*=4\Delta/\pi$ \cite{herbut2008coulomb,stauber2005disorder}. This shows the vital role of disorder in electronic properties of graphene which even can change the role of interaction and cause the system to exhibit the non-Frmi liquid behaviors \cite{kotov2012electron}. 

The polarization function of the graphene as an important quantity which is sensitive to both the inter-particle interaction and the disorder has been obtained and analyzed in the cone approximation in the bubble expansion~\cite{barlas2007chirality,hwang2007dielectric}. For the undoped graphene in one loop approximation of the polarization function (for which the intra-band excitations are forbidden due to the Pauli principle), it is shown that the static dielectric function $\epsilon(q)$ is a constant \cite{gonzalez1999marginal,kotov2012electron} and no collective modes are allowed within the PRA approximation. For finite $\mu$ however, the more general solution of Shung~\cite{Shung1986} is applicable which predicts that the potential is screened with the screening length $\lambda_{\text{TF}}\equiv \epsilon_0v_F/2e^2k_F$ ($v_F\equiv$ Fermi velocity, $k_F\equiv$ Fermi momentum, $\epsilon_0\equiv$ background dielectric constant). This predicts the Friedel oscillations in graphene to be of the form $\cos(2k_Fr)/k_Fr^3$~\cite{Shung1986}. Based on these results, in low-densities (in intrinsic or weakly disordered graphene) due to the lack of screening one should take into account the long-range part of the potential which leads to the issue of logarithmic phase shifts for Coulomb scatterers~\cite{kotov2012electron}. These results were obtained for disorder-free system and within the RPA approach which is questionable for the graphene as pointed out by Mishchenko \cite{mishchenko2007effect}. Despite of this huge literature, there is a limited understanding concerning the properties of the polarization function in the general form and also the behavior of multi-body charge density probability measures, especially in the low-density case in which EHPs appear.\\
A more reliable approach should involve both interaction and disorder simultaneously which is the aim of the present paper. For this, the Thomas-Fermi-Dirac (TFD) is a proper candidate which has the capacity to bring the effect of all LDA energies in the calculations as well as the tunable disorder in the same time. We employ this approach involving exchange-correlation energies and white noise out-plane disorder. For the EHPs, TFD techniques have proved to be very useful and are widely used \cite{adam2007self,hwang2007carrier,hwang2007density,rossi2008ground}. To obtain disorder corrections to the dielectric function we develop a diagramatic technique for the TFD theory. We obtain the two-body correlation function using some stochastic analysis and find the equation governing it. We show that, in addition to the screening length found by others, there is a characteristic length scale for the screening, namely $q_{\text{ch}}$ which is related to the disorder strength and inter-particle interactions via $q_{\text{ch}}\equiv \frac{\sqrt{\pi}}{4\sqrt{2}}dn_ir_s^2$ in which $n_i$ is the disorder strength, $r_s$ is the dimensionless interaction factor to be defined in the text and $d$ is the substrate distance. For small densities we show that there occurs an instability in which the imaginary part of the dielectric function becomes negative and some charge density oscillations with disorder-dependent wave length emerge. At these densities screening-anti-screening transition occurs along with changing sign of the real screened potential. The disorder significantly changes the Fourier component of the screened potential for large wave lengths showing the fact that the large-scale behaviors of the system is mostly affected by the disorder. We characterize the system in the vicinity of this transition in detail. In addition to the one- and the two-body  (static) correlation functions, the equation governing the total probability measure of the system is introduced and analyzed and a closed expression is proposed for it. The proposed expression is valid for large densities.\\
The paper has been organized as follows: In the next section we present the general construction of the problem and introduce the TFD theory for determining the ground state of graphene. The one-body probability measure is obtained in SEC. \ref{OneBody}. Section \ref{TwoBody} is devoted to finding the two-body probability measure and the dielectric function. The screened potential as well as the charge density oscillations are argued in this section. The total probability measure of the system is analyzed in SEC. \ref{TotalDistribution}. In the conclusion section we close the paper by highlighting the main findings of the paper. Appendix1 contains the essential rules for calculating Feynman diagrams for the TFD theory. Appendix2 and Appendix3 help to facilitate some calculations over the paper.

\section{General Construction; Ground state of Graphene}\label{GSG}

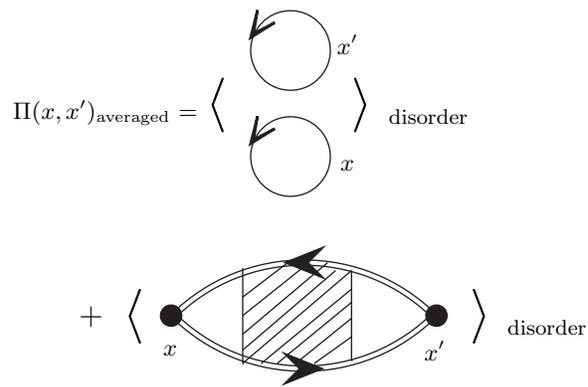
\begin{figure}
	\centering
	\begin{picture}(300,180)
	\SetArrowScale{2}
	\SetWidth{0.5}
	\SetColor{Black}
	\Text(40,140)[1]{ $\Pi(x,x')_{\text{averaged}}=$ \huge $ \left\langle \right. $}
	\Text(150,140)[1]{\huge $\left. \right\rangle$ \large $_{\text{disorder}} $}
	\ArrowArc[sep=2,arrow,arrowinset=0.8,arrowpos=0.4,arrowlength=10](105,120)(15,0,360)
	\ArrowArc[sep=2,arrow,arrowinset=0.8,arrowpos=0.4,arrowlength=10](105,160)(15,0,360)
	\Text(127,117)[1]{$x$}
	\Text(127,163)[1]{$x'$}
	\Text(30,60)[1]{\large +}
	\Text(45,60)[1]{ \huge $ \left\langle \right. $}
	\Text(195,60)[1]{\huge $\left. \right\rangle$ \large $_{\text{disorder}} $}
	\Text(60,47)[1]{$x$}
	\Text(160,47)[1]{$x'$}
	\ArrowArc[double,sep=2,arrow,arrowinset=0.4,arrowpos=0.5,arrowlength=13](110,10)(70,45,135)
	\ArrowArc[double,sep=2,arrow,arrowinset=0.4,arrowpos=0.5,arrowlength=13](110,110)(70,225,315)
	\CCirc(60,60){4}{Black}{Black}
	\CCirc(160,60){4}{Black}{Black}
	\Line(87,42)(87,78)
	\Line(87,42)(128,77)
	\Line(87,47)(122,77)
	\Line(87,53)(119,80)
	\Line(87,60)(112,80)
	\Line(87,67)(104,80)
	\Line(128,43)(128,77)
	\Line(94,42)(128,70)
	\Line(100,42)(128,65)
	\Line(107,42)(128,59)
	\Line(114,42)(128,52)
	\end{picture}
	\caption{(a) The diagrammatic representation of Eq \ref{Pi_Eq} for which the first term shows the connected component of the polarization operator, whereas the second term shows the disconnected component. Note that the disconnected term is also effectively connected via the external legs corresponding to the disorder which is averaged.}
	\label{fig:averaging}
\end{figure}

The effect of interaction and disorder in graphene has a long story in the literature. It is known that the random-phase approximation (RPA) fails to describe the interaction effects in graphene which was firstly pointed out by Mishchenko \cite{mishchenko2007effect}. The renormalization group analysis in the weak coupling limit in the first order approximation shows that the Coulomb interactions are marginally irrelevant due to the logarithmically divergent velocity~\cite{kotov2012electron}. In the strong interaction side however, it is shown that the graphene has non-Fermi liquid behaviors with power-law quasiparticle dispersion~\cite{son2007quantum,kotov2012electron}, the limit which cannot be reached for finite densities away from the Dirac point. In the weak coupling side, it is known that in contrast to ordinary metals in which the screening makes the interactions short-ranged, in low-density (intrinsic or weakly disordered) graphene due to the lack of screening one should take into account the long-range part of the potential which leads to the issue of logarithmic phase shifts for Coulomb scatterers~\cite{kotov2012electron}. The effect of disorder is however of special importance since, as stated in the introduction, it may completely change the behavior of the Dirac Fermions~\cite{herbut2008coulomb,stauber2005disorder}. TFD theory, as an approach which has the potential to bring the interaction and disorder in the calculations in the same footing is introduced and analyzed in this section. We mention some points concerning the TFD theory as a coarse grained method and the perturbation diagrams which are used in our analysis. The multi-point probability measures and their relations to the many body disordered response functions are of special importance in this paper which is described in the next sections.\\
Let us start by analyzing the zero-temperature polarization operator which is defined by $i\Pi(x,x')\equiv \hbar^{-1}\frac{\left\langle \Omega\right| T\left[\hat{n}(x)\hat{n}(x')\right] \left| \Omega\right\rangle }{\left\langle \Omega\right. \left| \Omega\right\rangle  } $ in which $T$ is the time ordering operator, $x\equiv (\textbf{r},t)$ is the space-time, $\hat{n}(x)\equiv \hat{\Psi}^{\dagger}(x)\hat{\Psi}(x)$ is the ground state density operator, $\hat{\Psi}$ and $\Psi^{\dagger}$ are the electron annihilation and creation operators and $\left| \Omega\right\rangle $ is the ground state of the system. Defining the density fluctuation operator as $\tilde{n}(x)\equiv \hat{n}(x)-\left\langle \Omega\right| \hat{n}(x) \left| \Omega\right\rangle$ (setting $\left\langle \Omega\right. \left| \Omega\right\rangle=1$) one finds:
\begin{equation}
\begin{split}
i\Pi(x-x')&=\left\langle i\Pi(x,x') \right\rangle_{\text{disorder}}\\
& =\hbar^{-1}\left\langle \left\langle \Omega\right|\hat{n}(x)\left| \Omega\right\rangle \left\langle \Omega\right|\hat{n}(x') \left| \Omega\right\rangle \right\rangle \\
& + \hbar^{-1}\left\langle \left\langle \Omega\right| T\left[\tilde{n}(x)\tilde{n}(x')\right] \left| \Omega\right\rangle\right\rangle
\end{split}
\label{Pi_Eq}
\end{equation}
in which the inner $\left\langle \right\rangle $ is the quantum expectation value, whereas the outer one stands for the disorder averaging. Note that although due to the presence of disorder this function (and other correlation functions) is not translational invariant, its disorder-averaged form is invariant. The first term is the disconnected part which has been shown in the first term in Fig. \ref{fig:averaging}, whereas the second part is connected term which has extensively been analyzed in the literature \cite{sarma2011electronic,kim2008graphene}. In this figure the single-line circles show the carrier density at the point $x$, i.e. $n(x)$ and $\left\langle \right\rangle_{\text{disorder}}$ shows the averaging over the disorder. In the same diagrams in Appendix1 the full-circles (double line circles) show the full density which is obtained self-consistently according to the employed theory, which is the TFD equations here. In that disorder-free systems the contribution of the first part is apparently trivial, since it generates an additional irrelevant constant. For the disordered systems however its effect is non-trivial as is extensively analyzed throughout this paper. In this paper we concentrate mainly on the contribution of this term and its effect on the response functions. As an example consider the ground state expectation value of the potential energy of the system. Let us define $n(x)\equiv\left\langle \Omega \left. \left| \hat{n}(x)\right| \right. \Omega\right\rangle$ which is the ground state density for a particular configuration of disorder and is not time-dependent (since $\left| \Omega \right\rangle $ has been chosen to be the exact ground state of the system), i.e. $n(x)=n(\textbf{r})$. Also let us define the second part as $G(\textbf{r}-\textbf{r}')\equiv \left\langle n(\textbf{r})n(\textbf{r}')\right\rangle_{\text{disorder}}$. The non-trivial effect of this term on this expectation value can be seen from the following relation \cite{fetter2012quantum} (the full treatment is postponed to Appendix1):
\begin{equation}
\begin{split}
\left\langle \left\langle \Omega\right| \hat{V}\left| \Omega\right\rangle \right\rangle|_{\text{first term}} = \frac{1}{2}\int d^2\textbf{r}d^2\textbf{r}' V(\textbf{r}-\textbf{r}')G(\textbf{r},\textbf{r}')
\end{split}
\label{average_potential}
\end{equation}
from which we see that the change of energy due to this term is $\delta E\sim \sum_qv_q^0G_q$ in which $v_q^0$ and $G_q$ are Furrier components of bare coulomb potential $v^0(r)$ and $G(r)$ respectively.\\
To investigate the effect of this term we turn to the TFD theory which contains coarse-graining of the space. In this case the contribution of the connected (second) term of the Fig. \ref{fig:averaging} becomes local, i.e. all the contributions are localized in a region in the close vicinity of the original spatial point, namely $x$. Therefore the exchange-correlation term becomes a local term in the Hamiltonian, and the only non-local terms are the Hartree and disorder terms. Appendix1 contains some diagrammatic analysis of the problem and the screened coulomb interaction as well as the spatial charge screening of the coulomb impurities have been analyzed. The Feynman diagrams need an especial care in this case since in the diagrammatic expansion of the energy (TFD energy), the only non-local terms are the Hartree and the external disorder interaction terms, as shown in Fig. \ref{fig:averaging1}. The coulomb potential is screened only via the mediator $G(\textbf{r}-\textbf{r}')$ as depicted in Fig. \ref{fig:averaging2} in which the full (double) lines involve simultaneously the disorder and coulomb interaction lines. The overall result is that considering such a mediator results in a change in the dielectric function $\delta\epsilon(q,\omega)\approx i\frac{\tilde{q}_0}{\tilde{q}}\tilde{G}(\tilde{q})$ in which $\tilde{G}$ is proportional to the Fourier component of $G(r)$, $\tilde{q}$ is proportional to $q$ (wave vector) and $\tilde{q}_0$ is a characteristic wave vector which has been defined in this appendix. All of these quantities will be defined and analyzed in the following sections. \\
The determination of $G(\textbf{r},\textbf{r}')$ need some information about the two-body distribution function. The multi-body distribution functions play an important role in the condensed matter physics, since the transport parameters are expressed in terms of these functions via the Kubo formulas. The other example in which we need such functions is the mean field theories in which the quantum degrees of freedom are reduced (the quantum fluctuations are killed) and the disorder averaging becomes the most important and challenging problem. The behavior of multi-point functions depend on the type and strength of the disorder.\\
For the more general case in which we are interested in calculating multi-point correlation functions with two insertion points ($\textbf{r}$ and $\textbf{r}'$), the extra contribution comes from the quantities like $\left\langle f(n(\textbf{r}),n(\textbf{r}'))\right\rangle $ (in which $f$ is an arbitrary function). In such cases we define the \textit{pair (or two-body) distribution measure} $P(n,\textbf{r};n',\textbf{r}')$ to write the mentioned correlation function as:
\begin{equation}
\begin{split}
\left\langle f(n(\textbf{r}),n(\textbf{r}'))\right\rangle =\int dndn'f(n,n')P(n,\textbf{r};n',\textbf{r}')
\end{split}
\end{equation}
for which $G(\textbf{r},\textbf{r}')$ is a special case. $P(n,\textbf{r};n',\textbf{r}')$ is defined as the probability density of charge density to be $n$ at $\textbf{r}$ and $n'$ at $\textbf{r}'$. It is notable that $P(n,\textbf{r};n',\textbf{r}')$ carries simultaneously the effect of the quantum system (Hamiltonian of the system) as well as the disorder statistics. Therefore finding the multi-point correlation functions reduces to finding multi-body probability densities. In the following we describe the TFD theory of the graphene monolayer to be used in the next sections as a quantum mechanical model to calculate these multi-body probability measures.\\
In graphene the carrier density is controlled by the gate voltage $n=\kappa_SV_g/4\pi t$ in which $\kappa_S$ is the substrate dielectric constant and $t$ is its thickness and $V_g$ is the gate voltage. The experimental data show a strong dependence on $x\equiv n/n_i$ in which $n_i$ is the impurity density. In ordinary densities, the conductivity is linear function of $x$ and for very low $x$'s, it reaches a minimum of order $\sigma\sim e^2/h$ which is linked with the formation of EHP's. Using local density approximation one can prove that the total energy of the graphene for a disorder configuration and a density profile is \cite{sarma2011electronic}:
\begin{eqnarray}
\begin{split}
E=&\hbar v_F[\frac{2\sqrt{\pi}}{3}\int d^2r\text{sgn}(n)|n|^{\frac{3}{2}}\\
&+\frac{r_s}{2}\int d^2r\int d^2r^{\prime}\frac{n(\textbf{r})n(\textbf{r}^{\prime})}{|\textbf{r}-\textbf{r}^{\prime}|}\\
&+r_s\int d^2rV_{xc}[n(\textbf{r})]n(\textbf{r})+r_s\int d^2rV_D(\textbf{r})n(\textbf{r})\\
&-\frac{\mu}{\hbar v_F}\int d^2rn(\textbf{r})]
\end{split}
\end{eqnarray}
in which $v_F$ is the Fermi velocity, $r_s\equiv e^2/\hbar v_F\kappa_S$ is the dimensionless interaction coupling constant, $\mu$ is the chemical potential, $g=g_sg_v=4$ is the total spin and valley degeneracy. The exchange-correlation potential is calculated to be \cite{polini2008density}:
\begin{eqnarray}
V_{xc}=\frac{1}{4}\left[ 1-gr_s\zeta(gr_s)\right]\text{sgn}(n)\sqrt{\pi|n|}\ln\left(4k_c/\sqrt{4\pi |n|}\right)
\end{eqnarray}
in which $k_c$ is the momentum cut-off and $\zeta(y)=\frac{1}{2}\int_0^{\infty}\frac{dx}{(1+x^2)^2\left( \sqrt{1+x^2}+\pi y/8\right)}$. The remote Coulomb disorder potential is calculated by the relation:
\begin{eqnarray}
V_D(r)=\int d^2r^{\prime}\frac{\rho(\textbf{r}^{\prime})}{\sqrt{|\textbf{r}-\textbf{r}^{\prime}|^2+d^2}}
\end{eqnarray}
in which $\rho(r)$ is the charged impurity density and $d$ is the distance between substrate and the graphene sheet. For the graphene on the SiO$_2$ substrate, $\kappa_S\simeq 2.5$, so that $r_s\simeq0.8$, $d\simeq 1$ nm, $k_c=1/a_0$ where $a_0$ is the graphene lattice constant $a_0\simeq 0.246$ nm corresponding to energy cut-off $E_c\simeq 3$ eV. 
One can easily minimize the energy with respect to $n(r)$ and obtain the following equation:
\begin{eqnarray}
\begin{split}
&\text{sgn}(n)\sqrt{|\pi n|}+\frac{r_s}{2}\int d^2 \textbf{r}^{\prime}\frac{n(\textbf{r}^{\prime})}{|\textbf{r}-\textbf{r}^{\prime}|}\\
&+r_sV_{xc}[n]+r_sV_D(\textbf{r})-\frac{\mu}{\hbar v_F}=0
\end{split}
\label{mainEQ}
\end{eqnarray}
which should be solved self-consistently. In this paper we consider the disorder to be white noise with Gaussian distribution $\left\langle \rho(\textbf{r})\right\rangle=0 $ and $\left\langle \rho(\textbf{r})\rho(\textbf{r}^{\prime})\right\rangle=(n_id)^2\delta^2(\textbf{r}-\textbf{r}^{\prime})$. \\
Let us now concentrate on the scaling properties of this equation excluding $V_{xc}$. By zooming out of the system, i.e. the transformation $\textbf{r}\rightarrow \lambda \textbf{r}$, we see that the equation remains unchanged if we transform $n(\textbf{r})\rightarrow n(\lambda\textbf{r})=\lambda^{-2}n(\textbf{r})$ as expected from the spatial dimension of $n(\textbf{r})$. This is because of the fact that $V_D(\lambda\textbf{r})=\lambda^{-1}V_D(\textbf{r})$. This symmetry is very important, since it causes the system to be self-affine. This scale-invariance in two dimension may lead to power-law behaviors and some exponents. It may also lead to conformal invariance of the system, and if independent of type of disorder, brings the graphene surface into a class in the minimal conformal series. The existence of $V_{xc}$ makes things difficult, since $V_{xc}(\textbf{r})\rightarrow V_{xc}(\lambda\textbf{r})=\lambda^{-1}\left(V_{xc}-\beta\text{sgn}(n)\sqrt{\pi|n|}\ln\lambda\right) $ in which $\beta\equiv\frac{1}{4}(1-gr_s\zeta(gr_s))$. Therefore the rescaled equation is:
\begin{eqnarray}
\begin{split}
&\xi(\lambda)\ \text{sgn}(n)\sqrt{|\pi n|}+\frac{r_s}{2}\int d^2 \textbf{r}^{\prime}\frac{n(\textbf{r}^{\prime})}{|\textbf{r}-\textbf{r}^{\prime}|}\\
&+r_sV_{xc}[n]+r_sV_D(\textbf{r})=0
\end{split}
\end{eqnarray}
in which $\xi(\lambda)\equiv 1-\beta r_s\ln\lambda$. Therefore the first term survive marginally in the infra-red limit and the scale invariance is expected, even in the presence of $V_{xc}$.

\section{One-Body probability density}\label{OneBody}

In this section we concentrate on calculating one-body probability density $P_n$ by focusing on Eq. \ref{mainEQ}. From now on $\left\langle \right\rangle $ stands for the disorder averaging. Using the identities $\nabla_{\textbf{r}}(\text{sgn}(n)\sqrt{\pi |n|})=\frac{1}{2}\text{sgn}(n)\sqrt{\frac{\pi}{|n|}}\nabla_{\textbf{r}} n$ and $\nabla_{\textbf{r}} V_{xc}=-\frac{1}{2}\beta\sqrt{\frac{\pi}{|n|}}[ 1-\text{sgn}(n)\ln( \frac{4k_c}{\sqrt{4\pi |n|}})] \nabla_{\textbf{r}} n$ we reach to the following identity:
\begin{eqnarray}
\begin{split}
& \nabla_{\textbf{r}} n=-r_sF_n\left\lbrace \overrightarrow{\chi}_{\rho}(d)+\frac{1}{2}\overrightarrow{\chi}_n(0)\right\rbrace 
\end{split}
\label{grad-n}
\end{eqnarray}
which is essential in the following sections. In this equation $F_n\equiv\frac{2\text{sgn}(n)\sqrt{|n|/\pi}}{1-r_s\beta\left[\text{sgn}(n)-\ln\frac{4k_c}{\sqrt{4\pi |n|}}\right]}$ and $\overrightarrow{\chi}_x(d)\equiv \int d^2r^{\prime}x(\textbf{r}^{\prime})\nabla \left(|\textbf{r}-\textbf{r}^{\prime}|^2+d^2\right)^{-1/2}$ and $x=\rho , n$. An important point is superficial contradiction that we have used $\nabla n$ which is ignored in the TFD theory, since it is valid only when $\frac{|\nabla n|}{n}\ll k_F$ ($k_F$ is the Fermi wave number). We turn back to this point at the end of this section.\\
The differential form of the charge profile is:
\begin{eqnarray}
\begin{split}
& dn=-r_sF_n\left[d\chi_{\rho}(d)+\frac{1}{2}d\chi_n(0)\right] 
\end{split}
\end{eqnarray}
Now let us perform some Ito calculations to obtain the probability measure of $n(r)$. We consider a one-body function $f(\textbf{r})=\left\langle f(n(\textbf{r}))\right\rangle$ and let $P(n,\textbf{r})$ be the probability measure of charge density to be $n$ at the position $\textbf{r}$, so that $f(\textbf{r})=\int dn P(n,\textbf{r})f(n)$. The differential of $f$ is defined as $df(\textbf{r})\equiv\left\langle f(n(\textbf{r})+dn)\right\rangle - \left\langle f(\textbf{r})\right\rangle$, according to which ($P_n(\textbf{r})\equiv P(n,\textbf{r})$):
\begin{eqnarray}
\begin{split}
&\text{d}f(\textbf{r})\equiv \int \text{d}P_n(\textbf{r})f(n)dn=\left\langle f(n(\textbf{r})+dn(\textbf{r}))-f(n(\textbf{r}))\right\rangle \\
&=\left\langle \partial_nf(n(\textbf{r}))dn(\textbf{r})+\frac{1}{2}\partial_n^2 f(\textbf{r})dn(\textbf{r})^2\right\rangle \\
& =\left\langle \left[ -r_sF_n\left( d\chi_{\rho}(d)+\frac{1}{2}d\chi_n(0)\right) \right]\partial_n f(n(\textbf{r}))\right\rangle \\
&+\left\langle \left[ \frac{r_s^2}{2}F_n^2(d\chi_{\rho})^2\right]\partial_n^2 f(n(\textbf{r}))\right\rangle 
\end{split}
\end{eqnarray}
Let us consider the above expression term by term. Firstly we note that $\left\langle \text{d}\chi_{\rho}=0\right\rangle$ which arises from the fact that $\rho(\textbf{r})=0$. The remaining two terms, can easily be transformed to the following form by using the integration by parts:
\begin{eqnarray}
\begin{split}
&df(\textbf{r}) =\int \text{d}P_n(\textbf{r})f(n)\\
&=\frac{r_s}{2}\int dn\left[\partial_n[F_nP_n(\textbf{r})\text{d}\chi_n]+ r_s\partial_n^2[f_n^2P_n(\textbf{r})(\text{d}\chi_{\rho})^2]\right]f(n) 
\end{split}
\label{Eq:oneBPM}
\end{eqnarray}
If we note that this equation should hold for all arbitrary $f$s, we can demand that the equality holds for the integral kernels and obtain:
\begin{eqnarray}
\begin{split}
dP(n,\textbf{r})=\frac{1}{2}r_s\partial_n\left[ F_nd\chi_nP_n\right] +\frac{r_s^2}{2}\partial_n^2\left[F_n^2(d\chi_{\rho})^2P_n\right] 
\end{split}
\end{eqnarray}
Now let us concentrate on $\left\langle (d\chi_{\rho})^2\right\rangle$ and $d\chi_n(0)$. The calculation of these quantities is done to appendix~\ref{appendix1} from which we see that $\left\langle (d\chi_{\rho})^2\right\rangle =\frac{\pi dn_i^2}{2\sqrt{2}}\text{d}r$, and also $d\chi_n(0)\simeq \frac{1}{\sqrt{2}}G_n \text{d}r$ in which $G_n=\int d^2\textbf{r}^{\prime}\frac{n(\textbf{r}^{\prime})}{|\textbf{r}-\textbf{r}^{\prime}|^2}$. If we substitute the above relations in the Eq.~\ref{Eq:oneBPM}, i.e. replacing $\text{d}\chi_n(0)$ and $\left( \text{d}\chi_{\rho}(d)\right)^2$ with their averages, we obtain:
\begin{eqnarray}
\begin{split}
\frac{\text{d}P(n,\textbf{r})}{\text{d}r}=\frac{r_s}{2\sqrt{2}}\partial_n\left[G_nF_nP_n\right] +\frac{\pi dn_i^2r_s^2}{4\sqrt{2}}\partial_n^2\left[F_n^2P_n\right] 
\end{split}
\end{eqnarray}
Therefore for the homogeneous system, which is independent of the observation point $\textbf{r}$, the left hand side of the above equation equals to zero which results to: 
\begin{eqnarray}
\begin{split}
\partial_n\left[G_nF_nP_n +\frac{1}{2}\pi dn_i^2r_s\partial_n\left[F_n^2P_n\right] \right]=0
\end{split}
\end{eqnarray}
By equating the expression inside the bracket to a constant (which is zero for the symmetry considerations) we obtain the equation governing the distribution of $n(r)$ as follows:
\begin{eqnarray}
\begin{split}
&\partial_n\left[F_n^2(d\chi_{\rho})^2P_n\right] =-\zeta_0 G_nF_nP_n\\
&\Rightarrow \partial_nP_n =- \Gamma_n P_n
\end{split}
\label{MainPnEq}
\end{eqnarray}
in which $\zeta_0\equiv \frac{2}{\pi dn_i^2r_s}$ and $\Gamma_n\equiv\zeta_0\frac{G_n}{F_n}+2\partial_n\ln F_n $. It is worth noting that in obtaining the above differential equation we have divided both sides of the equation by $F_n^2$ a task which is true only for non-zero finite $F_n$s. In fact $F_n$ is everywhere non-zero and well-defined and finite except at $n=0$ and $n\longrightarrow\infty$ for which becomes zero and negative infinity respectively. Therefore the equation~\ref{MainPnEq} is everywhere reliable except at $n=0$ for which, as we will see, becomes divergent and $n\longrightarrow\infty$ for which the solution is not reliable. This divergence at the Dirac point is due to the non-analytical behavior of $\nabla n|_{n\longrightarrow 0}$ in Eq.~\ref{grad-n}, i.e. the singular behavior of carrier density at the Dirac point. Such a behavior is missing in the other ordinary two-dimensional electronic systems and is specific to the Dirac electrons systems, since their kinetic energy densities are proportional to $\text{sgn}(n)|n|^{3/2}$ whose derivative is not well-behaved in $n=0$.\\
With the above limitation in mind, let us first turn to the asymptotic behaviors of Eq.~\ref{MainPnEq}. One may be interested in the answer for weak coupling limit $r_s\rightarrow 0$, or the weak disorder limit $n_i\rightarrow0$, i.e. large $\zeta_0$ limit. In this limit, and considering $G_n\approx G$ to be nearly constant, we have $\Gamma_n\simeq \frac{\sqrt{\pi}}{2}\zeta^{\prime}\frac{\text{sgn}(n)}{\sqrt{|n|}}=\zeta^{\prime}\partial_n\left(\text{sgn}(n)\sqrt{\pi|n|}\right)$ in which $\zeta^{\prime}\equiv \zeta_0 G$. The solution is therefore ($P_n(\textbf{r})=P_n$):
\begin{eqnarray}
\begin{split}
P_n^{\text{large $\zeta'$s}}=P_0\exp\left[-\zeta'\left( \text{sgn}(n)\sqrt{\pi |n|}-\frac{\mu}{\hbar S v_F}\right)\right]
\end{split}
\label{PnZeroth}
\end{eqnarray}
in which $S$ is the area of the sample and $P_0$ is the normalization constant. This relation may seem to be unsuited, since it grows unboundedly for negative values of $n$. Actually there is no contradiction, due to the presence of $G_n$ whose amount grows negatively for negative $n$ values, which returns the above equation to the expected form. In fact the original charge equation has electron-hole symmetry for the case $\mu=0$ which should result to an electron-hole symmetric form of $P_n$. Our approximation (considering $G_n$ as a constant) violated this symmetry. Re-considering this quantity as a dynamical variable retains the mentioned symmetry. It is also notable that the second term in the exponent ($\frac{\mu}{\hbar S v_F}$) has been inserted due to some symmetry considerations and the above equation satisfies the original equation of $P_n$.\\
In the above equation, the effects of disorder and interaction and $\mu$ have been actually coded in $\zeta^{\prime}$. Large amounts of $\zeta'^{-1}\sim \frac{r_s n_i^2}{G}$ results to a very large charge fluctuations, showing that the interaction and the disorder favor small deviations from mean density, whereas $G$ favors large charge fluctuations. To see how $G$ controls the charge fluctuations, we note that the limit $\mu\rightarrow 0$ (zero-gated graphene) has a direct effect on $G$. In fact $\mu$ controls $\left\langle n\right\rangle$ which directly affects $\int d^2\textbf{r}^{\prime}\ \frac{n(\textbf{r}^{\prime})}{|\textbf{r}-\textbf{r}^{\prime}|^2}$ which is $G$. In the limit $\mu\rightarrow 0$, we expect that $G$ becomes vanishingly small, so that ${\zeta^{\prime}}^{-1}\rightarrow \infty$ implying large scale density fluctuations, which in turn leads to EHPs.\\

\begin{figure*}
\centering
\begin{subfigure}{0.45\textwidth}\includegraphics[width=\textwidth]{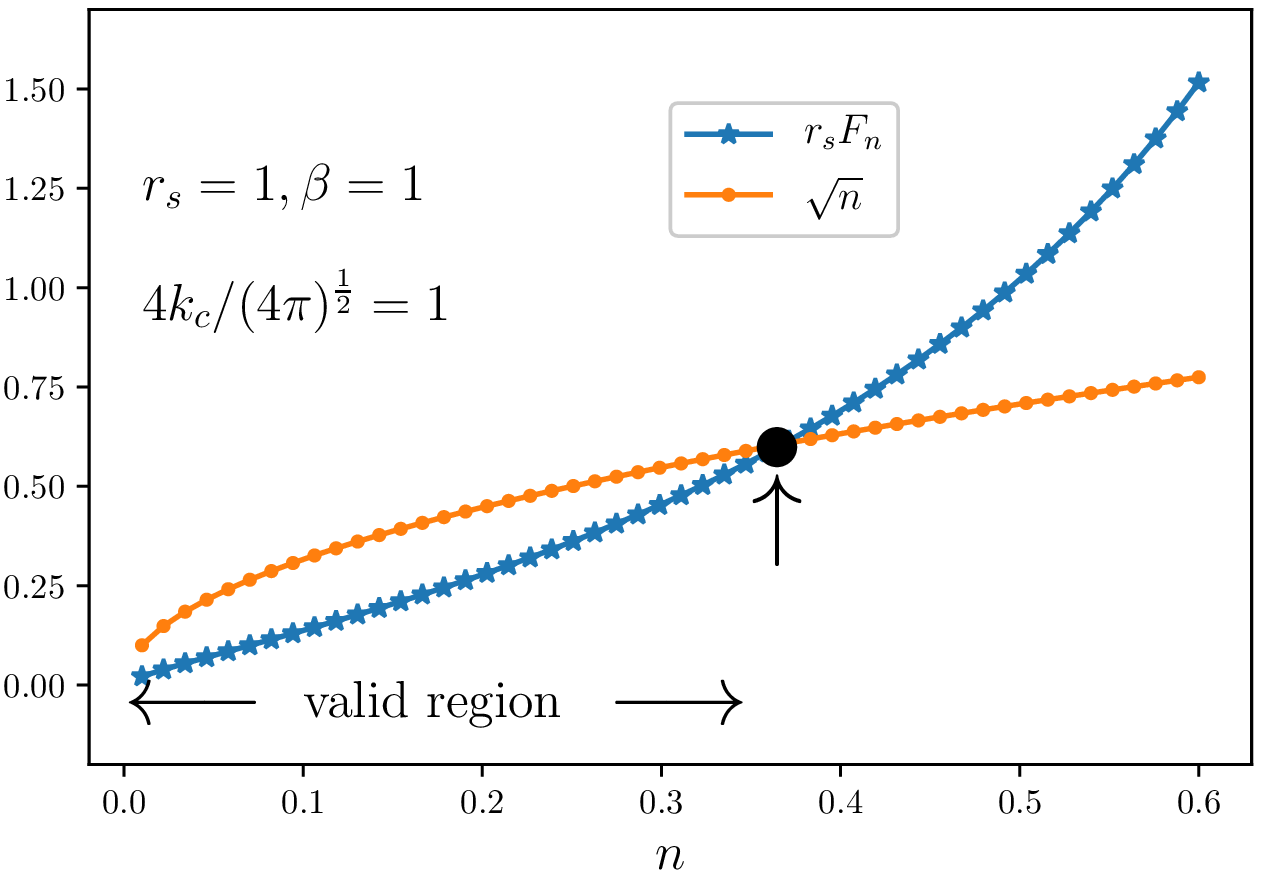}
\caption{}
\label{fig:validity}
\end{subfigure}
\begin{subfigure}{0.45\textwidth}\includegraphics[width=\textwidth]{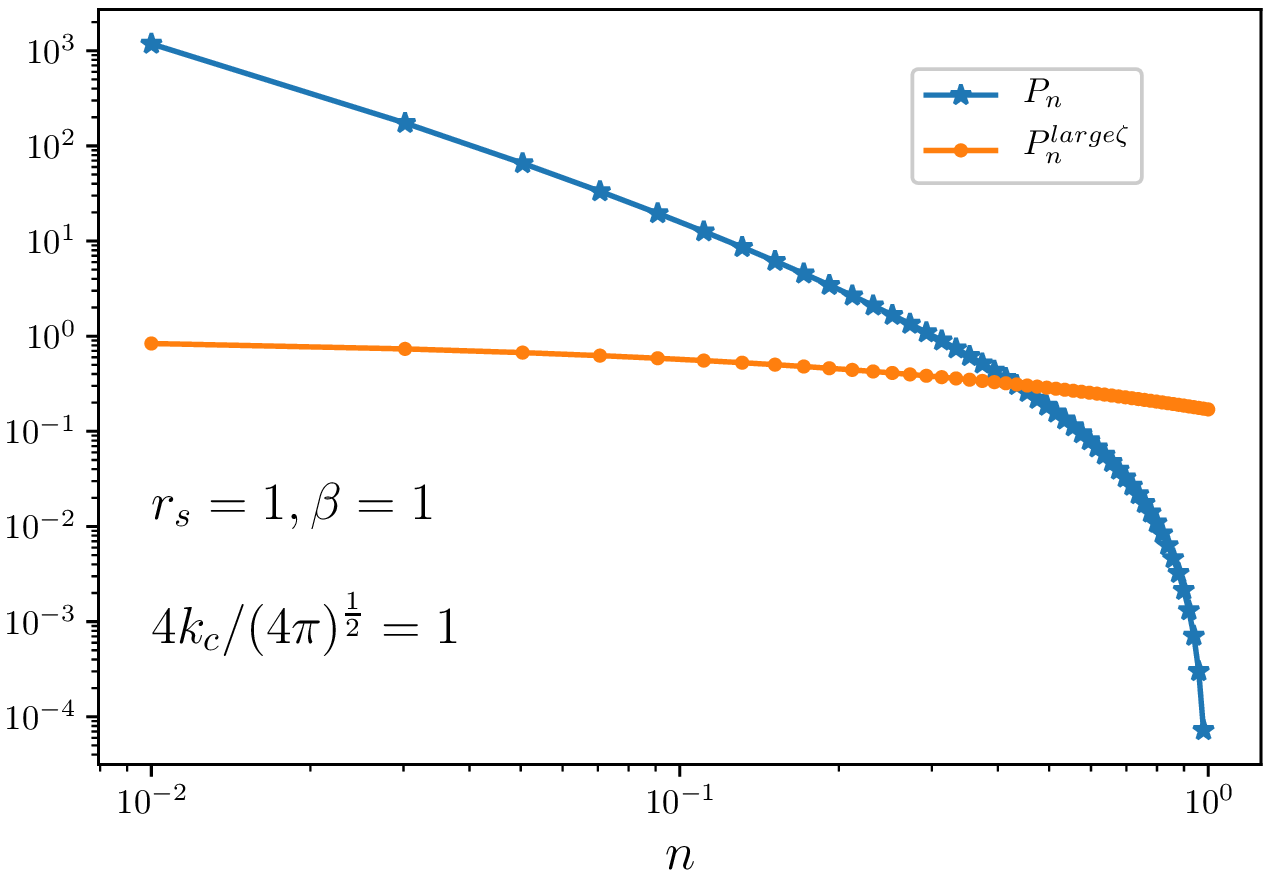}
\caption{}
\label{fig:Pn}
\end{subfigure}
\caption{(color online) (a) A comparison between $r_sF_n$ and $\sqrt{n}$ to see the validity region of the solution. (b) Log-log plot of the solutions $P_n^{\text{large $\zeta'$}}$ and $P_n$ in a valid range.}
\end{figure*}

The total solution of Eq~\ref{MainPnEq} is $P_n=P_n^{\text{large}\ \zeta'}f(n)$, in which :
\begin{eqnarray}
\begin{split}
f(n) & \equiv\frac{n^{\frac{1}{2}\zeta'\Omega\text{sgn}(n)\sqrt{\pi|n|}}}{|n|}\times\\
& \left(1-r_s\beta\left(\text{sgn}(n) -\ln \frac{4k_c}{\sqrt{4\pi |n|}}\right)\right)^2
\end{split}
\label{PnFullh}
\end{eqnarray}
In the above equation $P_n^{\text{large}\ \zeta'}$ is the solution \ref{PnZeroth}. As stated above, this solution is reliable for intermediate values of carrier density. To see the limitations of this solution, let us mention that the TFD solutions are valid only when $\frac{|\nabla n(\textbf{r})|}{n(\textbf{r})}\ll k_F(\textbf{r})=\sqrt{\pi |n(\textbf{r})|}$. Approximating $|\vec{\chi}_n|\sim G_n\sim n$, we find that $r_sF_n\ll n^{\frac{1}{2}}$. In Fig. \ref{fig:validity} we have shown a plot which compares $r_sF_n$ and $\sqrt{n}$ for $\beta=1$, $r_s=1$ and $\frac{4k_c}{\sqrt{4\pi}}=1$. We see that the validity of the solution \ref{PnFullh} is limited. Therefore for very high densities we do not expect that the one-body probability density be of the form \ref{PnFullh}. By the similar argument, we see that this solution is not the exact solution for $n\longrightarrow 0$ limit. In the Fig.~\ref{fig:Pn} we have shown $P_n^{\text{large $\zeta'$}}$ and $P_n$ for positive $n$s for $\mu=0$. Both of these functions have decreasing behavior in terms of the carrier density. In the case $\zeta'\longrightarrow\infty$, $P_n$ approaches continuously to $P_n^{\text{large $\zeta'$}}$ which admits larger carrier densities that is consistent with larger mean densities $\left\langle n\right\rangle$.
Before closing this section, let us turn to calculating the compressibility ($\kappa$). Knowing that the compressibility is proportional to $\left\langle n^2\right\rangle -\left\langle n\right\rangle^2$ we can easily do the integrals and show that, for the un-gated intrinsic graphene $\mu=0$, $\kappa\propto \zeta'^{-6}$. This shows that in the limit $\left\langle n\right\rangle \rightarrow 0$, $\kappa\rightarrow\infty$ (since $G\propto \left\langle n\right\rangle $) for which a transition is possible. Based on these observations, one may conclude that this transition is the formation of EHP landscapes.

\section{Two-body probability density}\label{TwoBody}

\begin{figure*}
\centering
\begin{subfigure}{0.47\textwidth}\includegraphics[width=\textwidth]{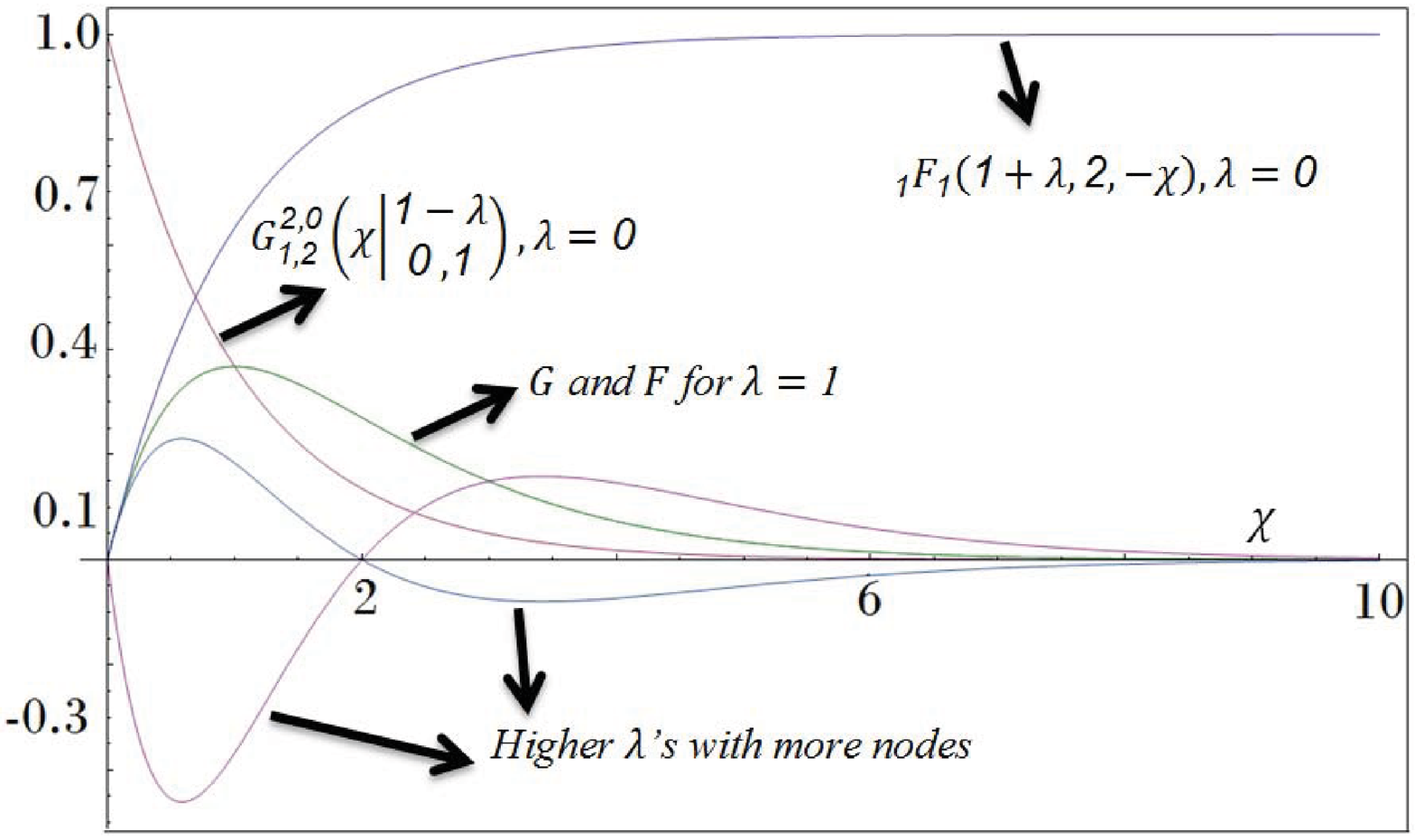}
\caption{}
\label{fig:Two-Body1}
\end{subfigure}
\begin{subfigure}{0.45\textwidth}\includegraphics[width=\textwidth]{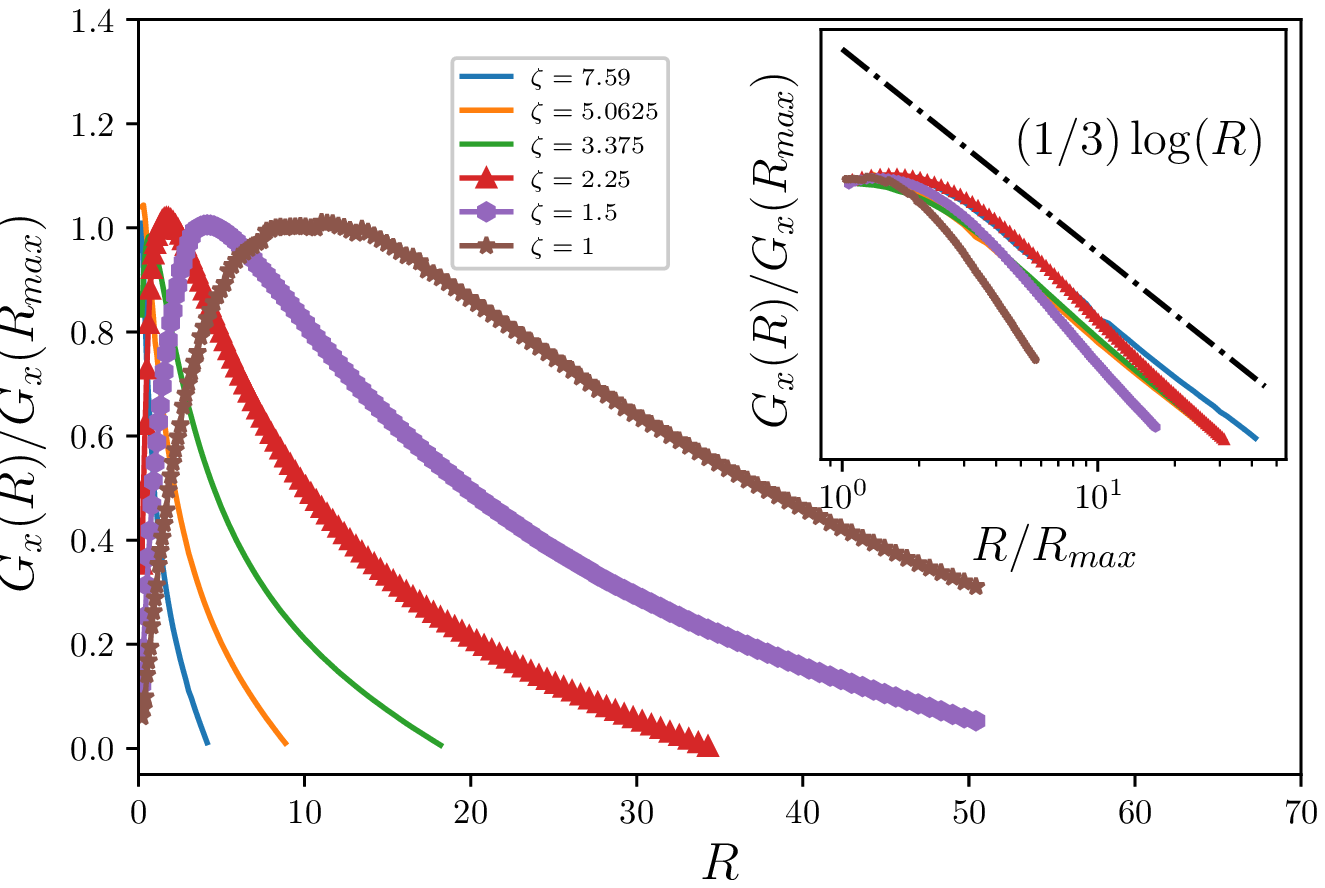}
\caption{}
\label{fig:G_r}
\end{subfigure}
\begin{subfigure}{0.45\textwidth}\includegraphics[width=\textwidth]{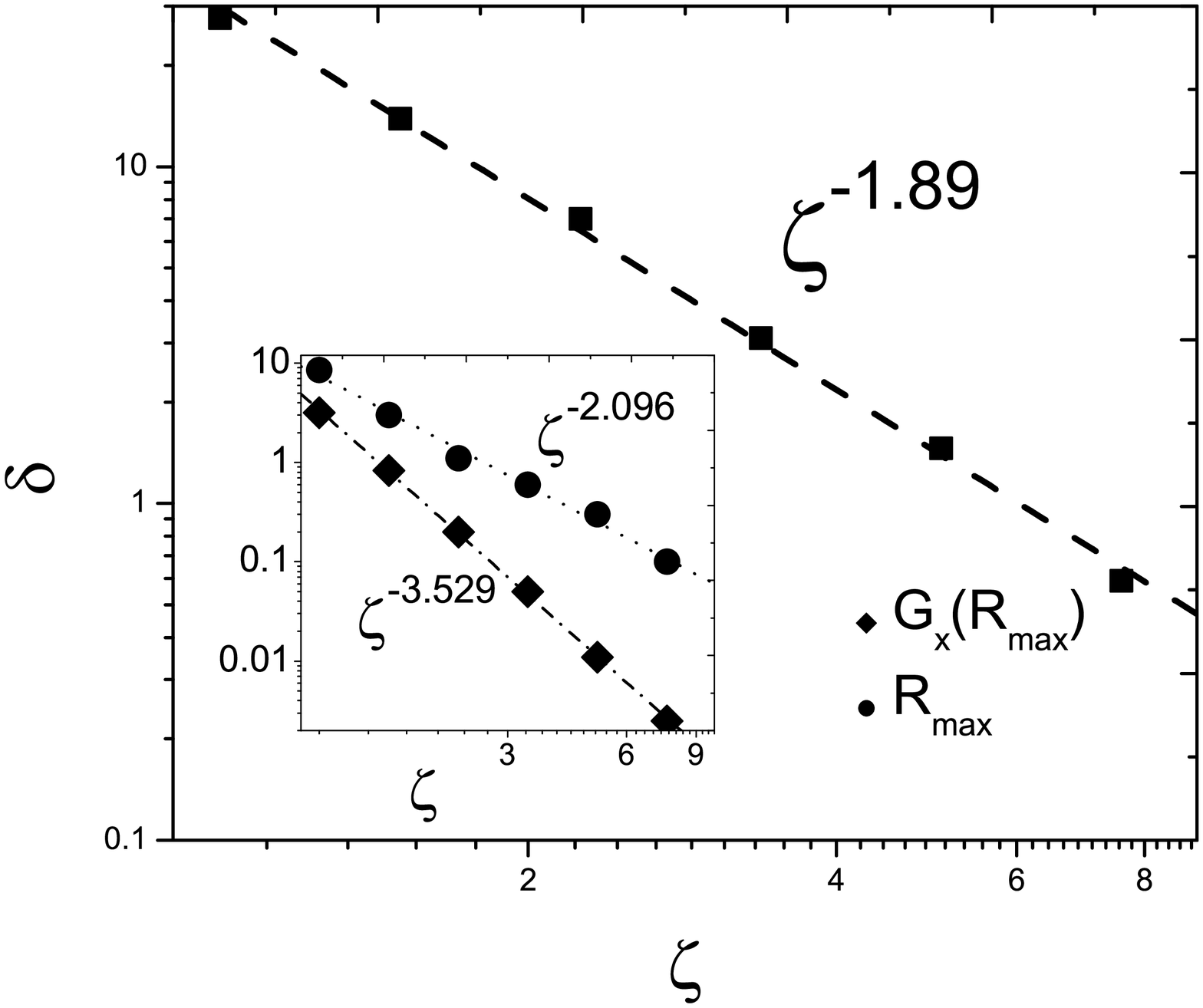}
	\caption{}
	\label{fig:Exponents}
\end{subfigure}
\caption{(Color online) (a) The plot of Kummer confluent hypergeometric and the Meijer G functions	for various $\lambda$'s. Higher $\lambda$ functions have some nodes which is forbidden. (b) The plot of $G_x(R)/G_x(R=0)$ in terms of $R$ for various rates of $\zeta$. Inset: The same plot in the semi-logarithmic form in which the $-\frac{1}{3}$ is evident. (c) The log-log plot of $\delta$ in terms of $\zeta$. Inset: The log-log plots of $G_x(R_{\text{max}})$ and $R_{\text{max}}$ in terms of $\zeta$. The power-law behaviors are evident in all of these graphs.}
\end{figure*}

In this section we consider a two-body functional. i. e. $f(\textbf{r},\textbf{r}')\equiv \left\langle f(n(r),n(r'))\right\rangle$. As stated in previous sections, $f(\textbf{r},\textbf{r}')=\int dndn'P(n,\textbf{r};n',\textbf{r}')f(n,n')$. The infinitesimal change of $f(\textbf{r},\textbf{r}')$ caused by the infinitesimal growth of the charge density is $df_{\textbf{r},\textbf{r}'}\equiv \left\langle f(n(\textbf{r})+dn,n(\textbf{r}'))\right\rangle-\left\langle f(n(\textbf{r}),n(\textbf{r}'))\right\rangle$. According to the previous section we have:
\begin{equation}
\begin{split}
&df_{\textbf{r},\textbf{r}'}=\\
&\left\langle \left[ -r_sF_n\left( d\chi_{\rho}(d)+\frac{1}{2}d\chi_n(0)\right) \right]\frac{\partial}{\partial n(\textbf{r})} f(n(\textbf{r}),n(\textbf{r}'))\right\rangle \\
&+\left\langle \left[ \frac{r_s^2}{2}F_n^2(d\chi_{\rho})^2\right]\frac{\partial^2}{\partial^2 n(\textbf{r})} f(n(\textbf{r}),n(\textbf{r}'))\right\rangle.
\end{split}
\end{equation}
Just like the calculations of the previous section we write (setting $\left\langle  d\chi_{\rho} \right\rangle =0$ and defining $n\equiv n(\textbf{r})$, $n'\equiv n(\textbf{r}')$, $P_{n,n'}(\textbf{r},\textbf{r}')\equiv P(n,\textbf{r};n',\textbf{r}')$):
\begin{equation}
\begin{split}
&df_{\textbf{r},\textbf{r}'}=\int \text{d}n\text{d}n' \\
&\left[ -r_sF_n d\chi_n\partial_n f_{n,n'}
+\frac{r_s^2}{2}F_n^2(d\chi_{\rho})^2\partial^2_n f_{n,n'}\right] P_{n,n'}(\textbf{r},\textbf{r}').
\end{split}
\end{equation}
from which, after iteration by parts and symmetrization and using also the definition $df_{\textbf{r},\textbf{r}'}\equiv \int\text{d}n\text{d}n' dP_{n,n'}(\textbf{r},\textbf{r}')f_{n,n'}$, we reach at the master equation for $P_{n,n'}$:
\begin{equation}
\frac{\text{d}}{\text{d}r}P_{n,n'}(\textbf{r},\textbf{r}')=\partial_n J_{n,n'}(\textbf{r},\textbf{r}')+\partial_{n'} J_{n',n}(\textbf{r},\textbf{r}')
\label{masterEq}
\end{equation}
in which:
\begin{equation}
\begin{split}
&J_{n,n'}(\textbf{r},\textbf{r}')\equiv\\
&\frac{r_s}{4}\left( F_nP_{n,n'}(\textbf{r},\textbf{r}')\frac{\text{d}\chi_n}{\text{d}r}+r_s\partial_n(F_n^2P_{n,n'}(\textbf{r},\textbf{r}'))\frac{(\text{d}\chi_{\rho})^2}{\text{d}r}\right).
\end{split}
\end{equation}
In the previous section we have calculated $\text{d}\chi_n$ and $(\text{d}\chi_{\rho})^2$ to be respectively $\frac{G_n}{\sqrt{2}}\text{d}r$ and $\frac{\pi dn_i^2}{2\sqrt{2}}\text{d}r$, leading to the relation ($P_{n,n'}\equiv P_{n,n'}(\textbf{r},\textbf{r}')$ and $J_{n,n'}\equiv J_{n,n'}(\textbf{r},\textbf{r}')$):
\begin{equation}
J_{n,n'}=\frac{r_s}{4\zeta_0\sqrt{2}}F_n^2\left[ \left( \zeta\frac{G_n}{F_n}+2\partial_n\ln F_n\right)P_{n,n'}+\partial_nP_{n,n'}\right].
\end{equation}
Let us consider the case $r_s\rightarrow 0$ or $n_i\rightarrow 0$, for which $F_n\approx \frac{2}{\sqrt{\pi}}\text{sgn}(n)\sqrt{n}$ and $G_n\approx$ constant. To facilitate the procedure we restrict the calculations to positive densities (so that $\text{sgn}(n)\sqrt{|n|}=\sqrt{n}$), having in mind that the same calculations should be done for negative densities (for which a minus sign is necessary). In this case $\zeta$ become very large and we have:
\begin{equation}
\begin{split}
\frac{\text{d}}{\text{d}r}P &=\alpha \left[ \partial_n(\sqrt{n}P)+\partial_{n'}(\sqrt{n'}P)\right]\\
&+ \gamma \left[\partial_n\left(n\partial_nP\right)+ \partial_{n'}\left(n'\partial_{n'}P\right)\right] 
\end{split}
\end{equation}
in which $\alpha\equiv r_sG/2\sqrt{2\pi}$ and $\gamma\equiv\frac{2}{\sqrt{\pi}}\frac{\alpha}{\zeta'}=\frac{r_s}{\zeta_0 \sqrt{2}\pi}=\frac{dn_i^2r_s^2}{2\sqrt{2}}$. We try the solution $P_{n,n'}=e^{-\zeta'(\sqrt{\pi n}+\sqrt{\pi n'})}P_0(r,n,n')$. Using this relation and the fact that $\sqrt{n}e^{-\zeta'\sqrt{\pi n}}P_0=-\frac{2}{\sqrt{\pi}}\frac{n}{\zeta'}\partial_n\left(e^{-\zeta'\sqrt{\pi n}}P_0\right)+\frac{2}{\sqrt{\pi}}\frac{n}{\zeta'}e^{-\zeta'\sqrt{\pi n}}\partial_nP_0$ we find that $P_0$ satisfies the following equation:
\begin{equation}
\partial_R P_0 =e^{\zeta\sqrt{\pi x}}\partial_x\left(e^{-\zeta\sqrt{\pi x}}x\partial_xP_0\right)+x\leftrightarrow x'.
\end{equation}
in which $R\equiv \frac{\alpha}{n_i\zeta'}r$, $x\equiv\frac{n}{n_i}$, $x'\equiv\frac{n'}{n_i}$ and $\zeta\equiv\zeta'\sqrt{n_i}$ are the dimensionless parameters. To solve this equation we consider $P_0$ to be composed of two parts, i.e. $P_0\equiv P_{0x}(\chi)P_{0x'}(\eta)$ in which $\chi\equiv \frac{x}{R}$ and $\eta\equiv\frac{x'}{R}$. Then we obtain two linear differential equations as follows:
\begin{equation}
\begin{split}
&\chi P_{0x}''+\left(1+\chi-\frac{\zeta\sqrt{\pi R}}{2}\sqrt{\chi}\right)P_{0x}'+\lambda P_{0x}=0\\
&\eta P_{0x'}''+\left(1-\eta+\frac{\zeta\sqrt{\pi R}}{2}\sqrt{\eta}\right)P_{0x'}'+\lambda P_{0x'}=0
\end{split}
\end{equation}
in which $P_{0x}''\equiv\partial_{\chi}^2P_{0x}$ and so on, and $\lambda$ is an arbitrary real number required in the separation of variables method. This form is not completely a factorized form since, as is seen the equations are not independent due to the presence of the common factor $R$. The second equation has been written in a form which is more suitable for our analysis. The initial form has been $\eta P_{0x}''+\left(1+\eta-\frac{\zeta\sqrt{\pi R}}{2}\sqrt{\eta}\right)P_{0x}'-\lambda P_{0x}=0$, and we have used the reflection symmetry $\eta\rightarrow -\eta$ (note that $\sqrt{\eta}$ has been $\text{sgn}(\eta)\sqrt{|\eta|}$ which changes sign under the mentioned operation). This means that we have two types of solution which is important in our analysis. For a while we suppose that $x>x'$ and seek for the solution of the above equation. Let us consider the limit of very small distances, namely $\alpha r\ll \sqrt{n}$ for which one can safely ignore $\zeta\sqrt{\chi}$ in the first equation and $\zeta\sqrt{\eta}$ in the second equation with respect to the other terms. The general solution of the equation $\chi P_{0x}''+\chi P_{0x}'+\lambda P_{0x}=0$ and the same for $P_{0x'}$ is hypergeometric and s:
\begin{equation}
\begin{split}
& P_{0x}(\chi)= A_{\lambda}\chi\  {_1}F_1(1+\lambda,2,-\chi) + B_{\lambda}G_{1,2}^{2,0}\left( \chi\left| \begin{array}{c} 1-\lambda \\  0\ ,\ 1 \end{array} \right. \right) \\
& P_{0x'}(\eta)= A'_{\lambda}\eta\ {_1}F_1(1-\lambda,2,\eta) + B'_{\lambda}G_{1,2}^{2,0}\left( -\eta\left| \begin{array}{c} 1+\lambda \\  0\ ,\ 1 \end{array} \right. \right)
\end{split}
\end{equation}
in which ${_1}F_1(a;b;-\chi)$ is the Kummer confluent hypergeometric function and $G_{p,q}^{m,n}\left( \chi\left| \begin{array}{c} a_1,...,a_p \\  b_1, ...,b_q \end{array} \right. \right)$ is the Meijer G function \cite{andrews1985special}, and $A_{\lambda}$, $B_{\lambda}$, $A'_{\lambda}$ and $B'_{\lambda}$ are some constants to be determined by the boundary conditions. For $\lambda=0, {_1}F_1(1+\lambda,2,-\chi)=\chi^{-1}(1-e^{-\chi})$ and $G_{1,2}^{2,0}\left( \chi\left| \begin{array}{c} 1-\lambda \\  0\ ,\ 1 \end{array} \right. \right)=e^{-\chi}$, and for $\lambda=1, {_1}F_1(1+\lambda,2,-\chi)=G_{1,2}^{2,0}\left( \chi\left| \begin{array}{c} 1-\lambda \\  0\ ,\ 1 \end{array} \right. \right)=\chi e^{-\chi}$. These functions for $\lambda=0,1$ and $2$ have been shown in Fig.~\ref{fig:Two-Body1}. To decide about the solutions we note that as $r\rightarrow \infty$ ($\chi\rightarrow 0$) the statistics of two disjoint parts become independent and we expect $P_{0x}(\chi\rightarrow 0)\rightarrow 1$. In the opposite limit, for $r\rightarrow 0$ ($\chi\rightarrow\infty$) $P_{n,n'}\rightarrow \delta(n-n')$, so that for $n \ne n'$, $P_{n,n'}\rightarrow 0$ in this limit, i.e. $P_{0x}(\chi\rightarrow\infty)\rightarrow 0$. These properties are only satisfied for $G_{1,2}^{2,0}\left( \chi\left| \begin{array}{c} 1-\lambda \\  0\ ,\ 1 \end{array} \right. \right)\left|  _{\lambda=0}\right. $, showing that all $A_{\lambda}=A'_{\lambda}=0$ for all $\lambda$'s and $B_{\lambda}=B'_{\lambda}=0$ for $\lambda \ne 0$. Noting that for $\lambda=0$ $G_{1,2}^{2,0}\left( \chi\left| \begin{array}{c} 1-\lambda \\  0\ ,\ 1 \end{array} \right. \right)=e^{-\chi}$ and $G_{1,2}^{2,0}\left( -\eta\left| \begin{array}{c} 1+\lambda \\  0\ ,\ 1 \end{array} \right. \right)=e^{\eta}$ we see that $P_0\propto \exp[-\frac{x-x'}{R}]$ for $x>x'$. The same solution can be done for the case $x'>x$ with interchanging the roles of $x$ and $x'$. This shows that $P_{n,n'}$ has the following important behavior:
\begin{equation}
P_{n,n'}=AP_x^{\text{large $\zeta$}}P_{x'}^{\text{large $\zeta$}} \exp\left[-\sqrt{\frac{2}{\pi}}\frac{4}{dn_ir_s^2}\frac{|x-x'|}{r}\right]
\label{Eq:Pnn}
\end{equation}
in which $A$ is some constant to be determined by normalization of $P$ and $P_x^{\text{large $\zeta$}}$ and  $P_{x'}^{\text{large $\zeta$}}$ are the solutions of equation \ref{PnZeroth}, i.e. one-body distribution functions. It is interestingly seen that for the limit $r\rightarrow\infty$ (which is beyond the approximation which was assumed above) this solution reduces to a factorized form of equation \ref{PnZeroth} which is expected since in this limit the particles behave as disjoint independent particles and the two-body distribution function should be multiplication of two one-body distribution function. The other important limit is $r\rightarrow 0$, for which the factor $e^{-\frac{x-x'}{R}}$ guarantees that in this limit $P_{n,n'}$ tends to zero except for the case $x=x'$ as expected.\\
A very important quantity is the density-density correlation function $G(r)=\left\langle n(r)n(0)\right\rangle$ which, as stated in the previous sections has very important information about the density response of the system to an external potential. It can easily be seen that the density-density correlation function is obtained to be $G(r)=n_i^2\left\langle x(r)x(0)\right\rangle\equiv n_i^2G_x(R)$, in which:
\begin{equation*}
\left\langle x(r)x(0)\right\rangle|_{\zeta} =\frac{\int_{-\infty}^{\infty}dx\int_{0}^{\infty}dUx(x-RU)I(\zeta,x,U,R)}{\int_{-\infty}^{\infty}dx\int_{0}^{\infty}dUI(\zeta,x,U,R)}
\end{equation*}
in which $I(\zeta,x,U,R)\equiv e^{-\zeta\left( \text{sgn}(x)\sqrt{\pi|x|}+\text{sgn}(x-RU)\sqrt{\pi|x-RU|}\right)}e^{-U}$ and we have used the new variable $U\equiv\frac{x-x'}{R}$. We have evaluated numerically this integral using the iterative integration method with summation steps $\delta U=\delta x=0.001$ and precision ratio $10^{-5}$ in each direction (one for the $U$-direction and another for $n$-direction). The result has been shown in Fig. \ref{fig:G_r} in which $\frac{G_x(R)}{G_x(R_{\text{max}})}$ has been sketched for various rates of $\zeta$ ($R_{\text{max}}$ is the amount of $R$ in which $G_x(R)$ shows a peak). It can be shown that $G_x(R=0)\sim \zeta^{-6}$ as stated in the end of the previous section. This function tends to zero as $R\rightarrow\infty$ as expected. $G_x(R)$ has some interesting features. Firstly the tail of this function (for large $R$'s) behaves in a logarithmic fashion. To show this we have shown $G_x(R)/G_x(R_{\text{mean}})$ in terms of re-scaled $R/R_{\text{mean}}$ in a semi-logarithmic graph in the inset of Fig. \ref{fig:G_r} for which a $-\frac{1}{3}$ slope is evident. Except for small $\zeta$s ($\zeta=1$ in this graph), the graphs fit properly to each other with a nearly the same slope. Clearly this behavior can't remain for very large $R$s, since $G_x(R)$ should be positive in all $R$ range. The logarithmic behavior in graphene has been observed in many aspects, e.g. the logarithmic enhancement of Fermi velocity \cite{gonzalez1999marginal}, or exchange-driven singularity in the Fermi velocity in intrinsic graphene \cite{vafek2007anomalous}, but up to the author's knowledge such a behavior in two-point correlation functions has not been reported before. The logarithmic behavior is the characteristics of \textit{free Boson} systems that our effective model for intermediate spatial scales corresponds to. The other feature of $G_x(R)$ is the narrowing of its peak as $\zeta$ increases, and that $R_{\text{max}}$ and $G_x(R_{\text{max}})$ are decreasing functions of $\zeta$. The change of these quantities has been shown in the Fig. \ref{fig:Exponents}. It is interestingly seen that these quantities show some clean power-law behaviors in terms of $\zeta$. The power-law behavior arises from the scale-invariance of the TFD equation which was analyzed at the end of SEC. \ref{GSG}. To characterize the behavior of $G_x(R)$ let us analyze the width of the $G_x(R)$ defined by $\delta\equiv R_2-R_1$ in which $G_x(R_1)=G_x(R_2)=e^{-1}G_x(R_{\text{max}})$, and also $G_x(R_{\text{max}})$ and $R_{\text{max}}$ in terms of $\zeta$. The behaviors are $\delta\sim \zeta^{-\tau_{\delta}}$, $G_x(R_{\text{max}})\sim \zeta^{-\tau_{G}}$ and $R_{\text{max}}\sim \zeta^{-\tau_R}$ in which $\tau_{\delta}\approx 1.89$, $\tau_G\approx 2.1$ and $\tau_R\approx 3.6\approx 2\tau_{\delta}$, which arises to the hyper-scaling relation:
\begin{equation}
\delta\sim R_{\text{max}}^{\frac{1}{2}}.
\end{equation}
In the limit in which the interaction is strong enough and (or) the fraction $\left\langle x\right\rangle$ is small enough, i.e. $\zeta$ is small, $R_{\text{max}}$ becomes large and therefore $\delta$ increases unboundedly which leads to large density inhomogeneity, i.e. formation of EHPs.\\

\begin{figure*}
	\centering
	\begin{subfigure}{0.45\textwidth}\includegraphics[width=\textwidth]{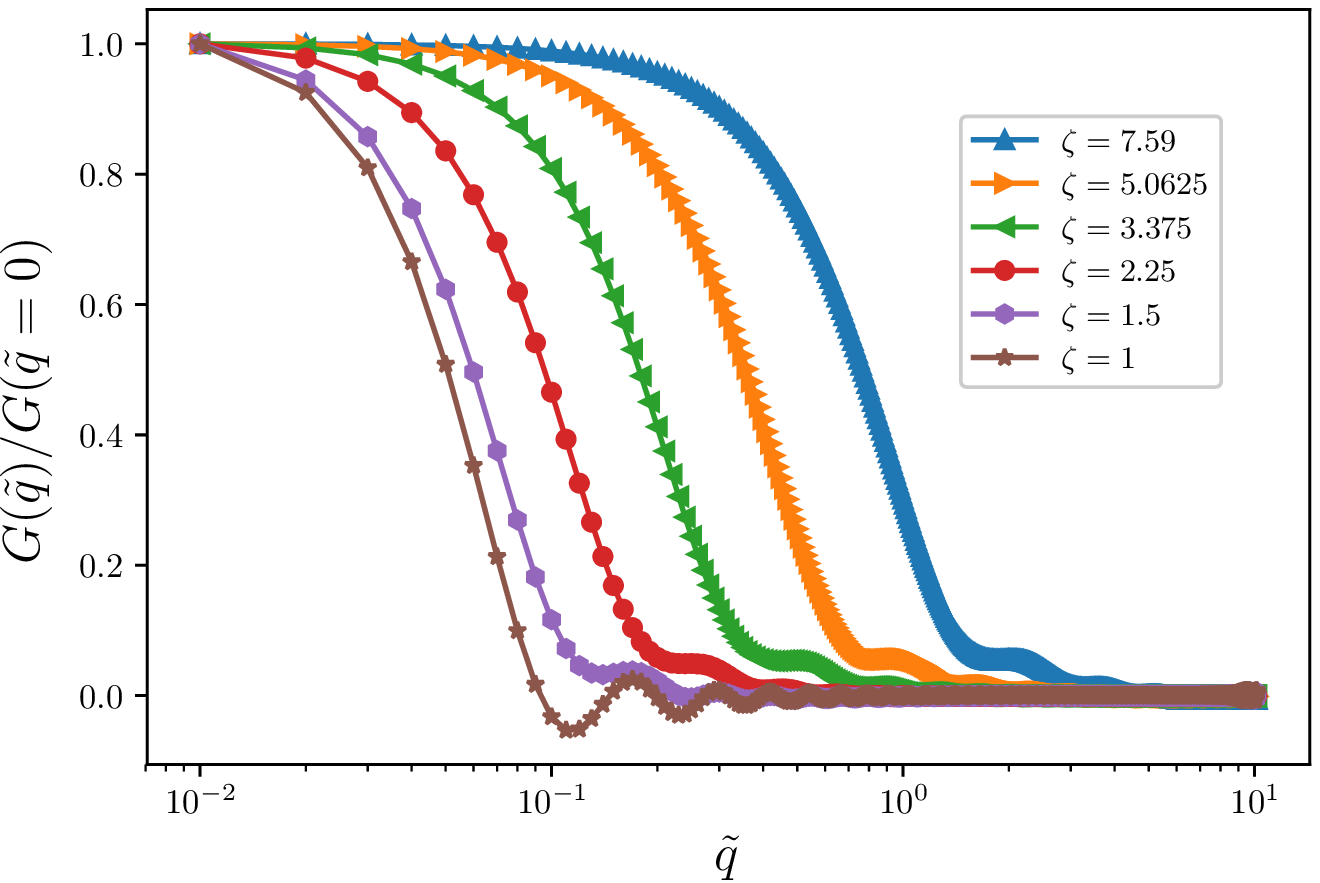}
		\caption{}
		\label{fig:G_q}
	\end{subfigure}
	\begin{subfigure}{0.45\textwidth}\includegraphics[width=\textwidth]{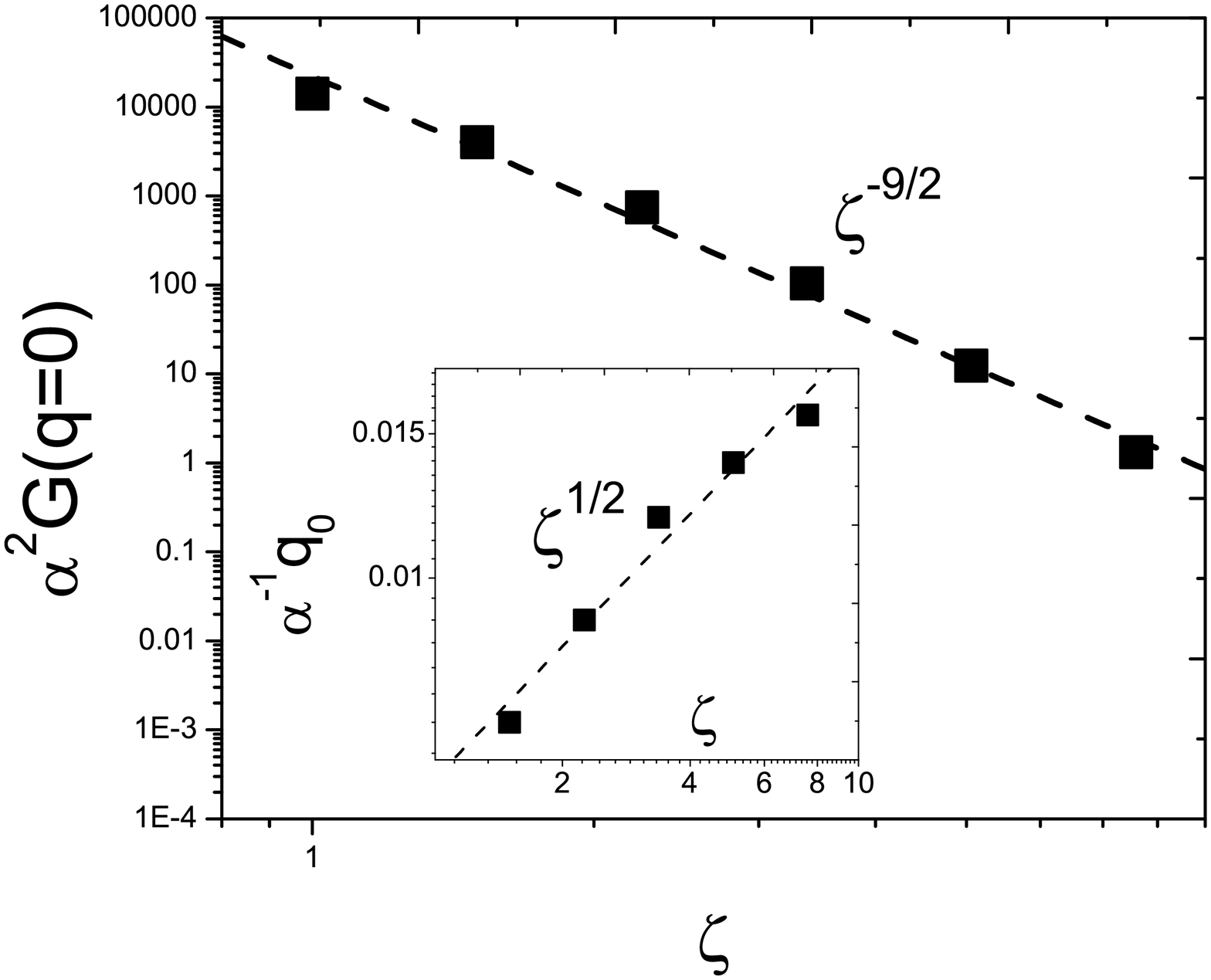}
		\caption{}
		\label{fig:G_q=0}
	\end{subfigure}
	\begin{subfigure}{0.45\textwidth}\includegraphics[width=\textwidth]{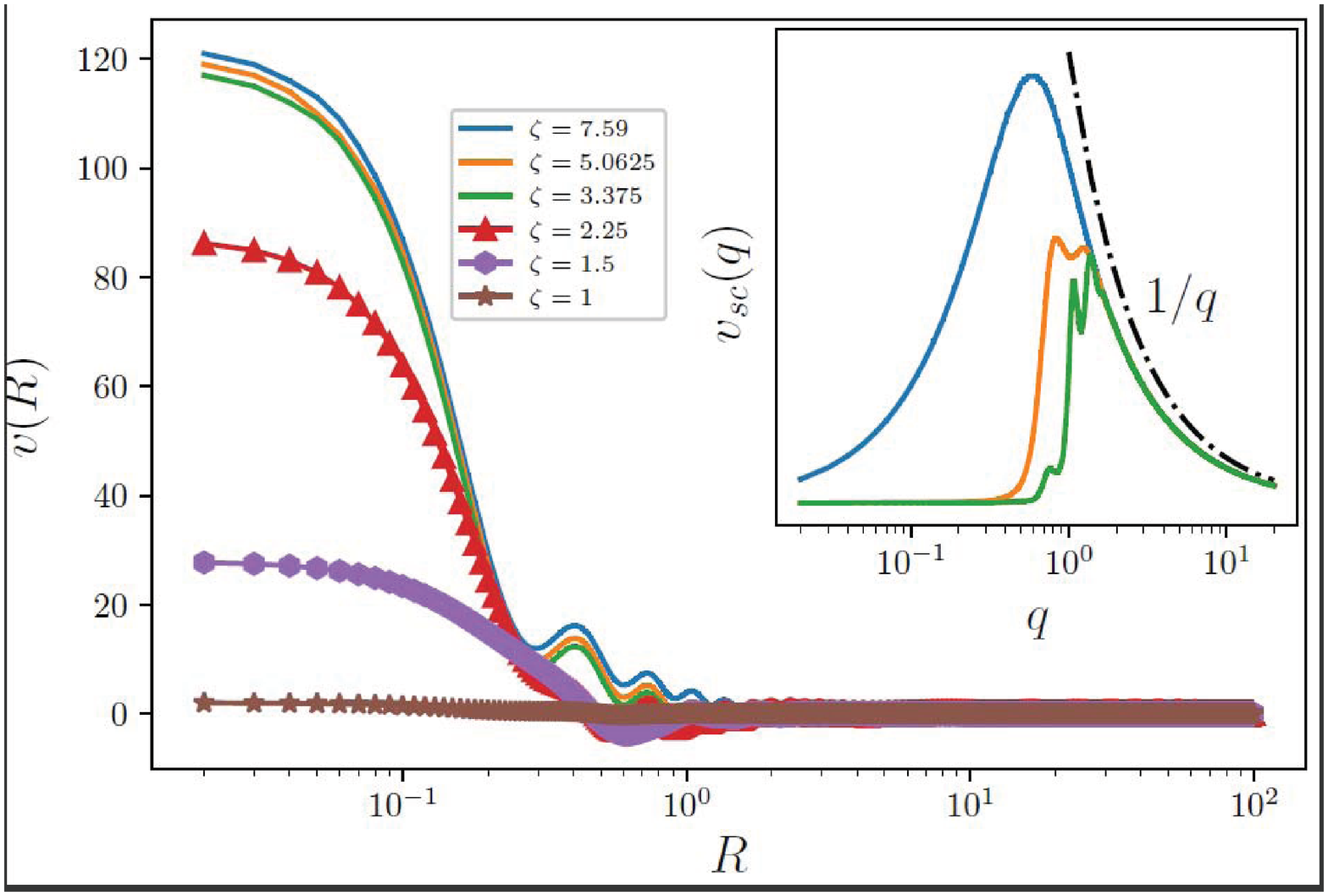}
		\caption{}
		\label{fig:screenedPotential}
	\end{subfigure}
	\begin{subfigure}{0.45\textwidth}\includegraphics[width=\textwidth]{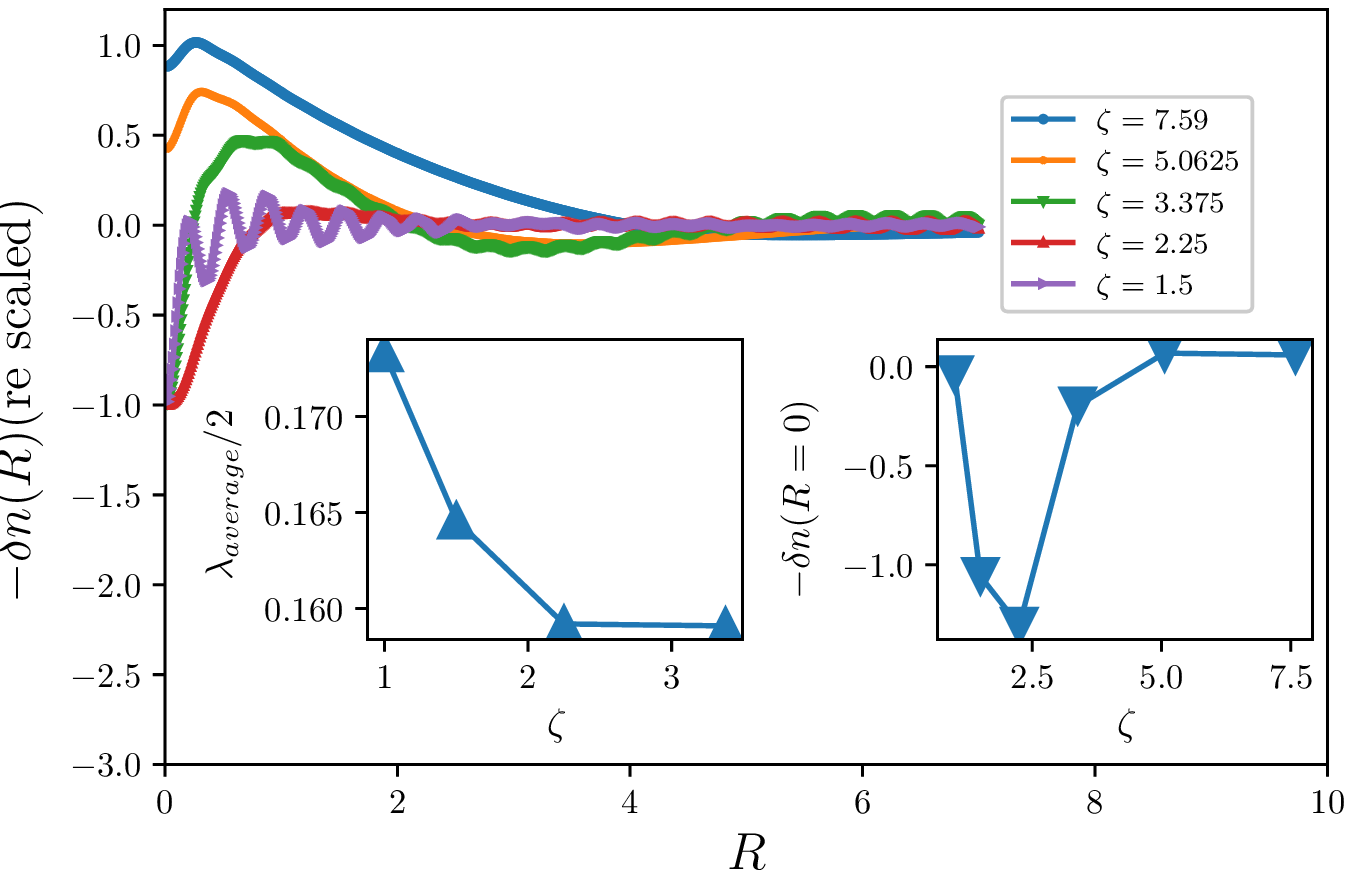}
		\caption{}
		\label{fig:delta_n_r}
	\end{subfigure}
	\caption{(color online) (a) The semi-log plot of $\tilde{G}(\tilde{q})/\tilde{G}(\tilde{q}=0)$. In this graph $q_0$ is defined as the point above which the graph falls off rapidly. (b) The power-law behavior of $\alpha^2G(q=0)$ in terms of $\zeta$ with the exponent $\frac{9}{2}$. Inset: The log-log plot of $\alpha^{-1}q_0$ in terms of $\zeta$ with the exponent $\frac{1}{2}$. (c) The real-space screened potential which is deformed significantly for smaller $\zeta$'s. For $\zeta\lesssim 3$ the potential becomes negative for some $R$ intervals which is the finger print of some instability. Inset: The semi-log plot of the Fourier component of the screened potential. The potential if deformed most significantly for small $q$'s and is magnified for smaller $\zeta$'s, showing that this deformation is disorder-driven. (d) The real-space screening of charge impurity and its oscillatory behavior for small $\zeta$'s. The right inset shows the screening at $R=0$. It is seen that for $\zeta\simeq 3$ changes sign which signals the screening-anti-screening transition. The average wave length of the oscillatory behavior has been shown in the left inset.}
\end{figure*}

The other important quantity is the Fourier transform of $G_x(R)$, i.e. $G(q)$ which is shown to be:
\begin{equation}
\begin{split}
\text{Re}\left[ G(q)\right] & =2\pi n_i^2 \left( \frac{\zeta'n_i}{\alpha}\right)^2\int_0^{\infty}dRRJ_0\left( \tilde{q}R\right)G_x(R)\\
& = \frac{32}{\pi d^2r_s^4}\tilde{G}(\tilde{q})
\end{split}
\label{G_x}
\end{equation}
in which $J_n(x)$ is the Bessel function of the first kind, $\tilde{q}\equiv \frac{n_i\zeta'}{\alpha}q$ and $\tilde{G}(\tilde{q})\equiv 2\pi \int_0^{\infty}dRRJ_0\left( \tilde{q}R\right)G_x(R)$. Note that the imaginary part of $G(q)$ vanishes. The presence of disorder induces a characteristic wave vector $q_{\text{ch}}\equiv \frac{\alpha}{n_i\zeta'}$ (so that $\tilde{q}=\frac{q}{q_{\text{ch}}}$), which implies that $\frac{q_{\text{ch}}}{k_F}\approx \sqrt{\frac{G}{2}}\frac{r_s}{2\pi n_i\zeta'}$. In this relation we have used $k_F=\sqrt{\pi \left\langle n\right\rangle }$ and $G\approx \left\langle n\right\rangle$. $\text{Re}\left[ \tilde{G}(\tilde{q})\right]/\tilde{G}(\tilde{q}=0) $ has been sketched in Fig. \ref{fig:G_q} for various rates of $\zeta$ in a semi-logarithmic scale. For large $\zeta$ values this function is nearly constant for small values of $\tilde{q}$ and starts to falling off rapidly at some $q$ value which we name $q_0$. For $q$ values in the order of (and larger than) $q_{\text{ch}}$, we enter the phase in which $\text{Re}\left[ G(q)\right]/G(q=0) $ is vanishingly small. $q_0$ is $\zeta$-dependent and increases as $\zeta$ increases. It is interestingly seen that $G(q=0)$ and $q_0$ show power-law behaviors in terms of $\zeta$ with exponents shown in the figure, i.e. $\alpha^2\tilde{G}(q=0)\sim \zeta^{-\tau_{\tilde{G}}}$ and $\alpha^{-1}q_0\sim \zeta^{\tau_{q}}$ in which $\tau_{\tilde{G}}\approx 9/2$ and $\tau_q\approx 1/2$. \\
To understand where these power-law behaviors come from, we should look at the main equation of $n(\textbf{r})$, i.e. Eq. \ref{mainEQ}. This behavior has its roots in the scale invariance of this equation (to see this symmetry let $\textbf{r}\rightarrow \lambda \textbf{r}$ and $n(\textbf{r})\rightarrow \lambda^{-2}n(\textbf{r})$ which yields the same equation with a renormalized logarithmic-enhanced coefficient) which cause a scale symmetry in Eq. \ref{Eq:Pnn}. In fact if we do the transformations $\zeta\rightarrow \lambda^{-1}\zeta$, $x\rightarrow \lambda^2 x$ and $R\rightarrow \lambda^2 R$, then we see that $G_x(r)\rightarrow \lambda^4 G_x(r)$. \\
The effect of $G(q)$ on the screened potential is very interesting. The Fourier component of the screened potential ($v_{\text{sc}}(q)$) in long wave lengths (small $q$ with respect to $q_{\text{ch}}$) is crucially changed, so that $v_{\text{sc}}(q\rightarrow 0)\rightarrow 0$, whereas the form of $v_{\text{sc}}$ is not changed with respect to the bare interaction potential for large and intermediate wave numbers. It has been shown in the inset of Fig. \ref{fig:screenedPotential} from which it is seen that all curves are fitted to $1/q$ for large $q$'s. The cross-over between these two behaviors occur in some $q$ interval. The potential in the real space $v(R)$ has some interesting features. For large $R$'s and large $\zeta$'s this quantity falls off just like the bare potential, i.e. $1/R$ with some oscillations which is evident in Fig. \ref{fig:screenedPotential}. In the small $R$ limit the potential tends to a constant value, which cause $v(R=0)$ to be finite. When $\zeta$ decreases ($\zeta\lesssim 3.5$) the behaviors change crucially as is seen in the figure and becomes negative in some $R$ intervals, signaling the instability of the Dirac gas. This change of behavior can be seen in the response function $\delta n(R)\equiv -\int d^2q v_{\text{sc}}(q)G(q)$ which has been shown in Fig. \ref{fig:delta_n_r}. In the right inset $-\delta n(R=0)$ has been shown in terms of $\zeta$, from which a screening-anti-screening transition is observed for $\zeta\sim 3.5$. This corresponds to the change of potential sign, as explained above. An oscillatory behavior appears with the wave length $\lambda_{\text{average}}$ shown in the left inst of this figure for small $\zeta's$. The corresponding wave numbers are consistent with the ones for which the potential $v_{\text{sc}}(q)$ shows a peak. The emergent oscillatory behavior is the effect of characteristic wave number for which the screened potential shows peak.\\
Before closing this section, it is worth mentioning some points concerning the form of $G(q)$. It is well-known that for zero temperature case the polarization given by the RPA for mono-layer garphene is constant for $q<2k_F$ and increases nearly linearly for $q>2k_F$. This is in sharp contrast to the ordinary two-dimensional electron gas (2DEG) in which for $q>2k_F$, it is a decreasing function of $q$ with a discontinuity in the derivative at $q=2k_F$ \cite{sarma2011electronic}. $G(q)$ (as the contribution of the second term in Fig. \ref{fig:averaging}) shows actually an opposite behavior with the other reference point $q_{\text{ch}}$ which is more similar to the behavior of ordinary 2DEG. Therefore one may be interested in the case in which the second term dominates the first term in Fig. \ref{fig:averaging}, which cause the graphene to behave like ordinary 2DEG. This may be possible for small enough $\zeta$s (strong interactions, strong disorder and small densities) for which $G(q=0)$ is larger and electron-hole puddles are present. The total information concerning the two-body charge density distribution function has been gathered in TABLE. \ref{tab:exponents}.

\begin{table}
	\begin{tabular}{c|c|c}
		\hline Exponent & Definition & closest fractional amount \\
		\hline $a_G$ & $G_x(R)\propto a_G\log(R)$ & $\frac{1}{3}$ \\
		\hline $\tau_G$ & $G_x(R_{\text{max}})\propto \zeta^{-\tau_G}$ & $\frac{7}{2}$ \\
		\hline $\tau_R$ & $R_{\text{max}}\propto \zeta^{-\tau_R}$ & $2$ \\
		\hline $\tau_{\delta}$ & $\delta\propto\zeta^{-\tau_{\delta}}$ & $\tau_G/2$ \\
		\hline $\tau_{\tilde{G}}$ & $\alpha^2\tilde{G}(q=0)\propto \zeta^{-\tau_{\tilde{G}}}$ & $\frac{9}{2}$ \\
		\hline $\tau_{q}$ & $\alpha^{-1}q_0\propto \zeta^{\tau_{q}}$ & $\frac{1}{2}$ \\
		\hline $\gamma_{\delta,R}$ & $\delta\propto R_{\text{max}}^{\gamma_{\delta,R}}$ & $\frac{1}{2}$ \\
		\hline
	\end{tabular}
	\caption{The critical exponents of graphene related to the two-body charge density distribution function with their definitions.}
	\label{tab:exponents}
\end{table}

\section{The charge probability measure}\label{TotalDistribution}

In this section we seek for the equation governing $P\left(\left\lbrace n\right\rbrace\right)$ which is the probability measure for the charge configuration $\left\lbrace n\right\rbrace$. This function actually depends not only to $n$ configuration, but also to its gradient: $P\left( \left\lbrace n\right\rbrace \right)\equiv P\left(\left\lbrace n\right\rbrace , \left\lbrace \nabla n\right\rbrace \right)$. But as we will see the dependence to the local charges is enough when we are looking at the large carrier densities. Let us suppose that $f$ is an arbitrary local or non-local smooth function of the density $n$, i.e. $f_{\textbf{r}_0}=\int d^2\textbf{r}f(n(\textbf{r}),n(\textbf{r}_0))g(\textbf{r},\textbf{r}_0)$ in which $\textbf{r}_0$ is some reference point. Without loose of generality we set $g\equiv1$ to facilitate the calculations, having in mind that there is some reference point from which the positions are calculated. Using the the equations of SEC~\ref{GSG}, it can therefore be expanded in terms $n$ (defining $f_{\textbf{r}_0}(n(\textbf{r}))\equiv f(n(\textbf{r}),n(\textbf{r}_0))$ and $dn(\textbf{r})=n(\textbf{r}+d\textbf{r})-n(\textbf{r})$): 
\begin{eqnarray}
\begin{split}
&df_{\textbf{r}_0}(n)\equiv \int d^2\textbf{r}[f_{\textbf{r}_0}(n(\textbf{r})+dn(\textbf{r}))-f_{\textbf{r}_0}(n(\textbf{r}))]\\
&\equiv \int d^2\textbf{r}[f_{\textbf{r}_0-d\textbf{r}}(n(\textbf{r}))-f_{\textbf{r}_0}(n(\textbf{r}))]\\
&=\int d^2\textbf{r}[\partial_nf_{\textbf{r}_0}(n(\textbf{r}))dn(\textbf{r})+\frac{1}{2}\partial_n^2 f_{\textbf{r}_0}(\textbf{r})dn(\textbf{r})^2]\\
& =\int d^2\textbf{r}\left[ -r_sF_n\left( d\chi_{\rho}(d)+\frac{1}{2}d\chi_n(0)\right) \right]\partial_n f_{\textbf{r}_0}(n(\textbf{r}))\\
&+\int d^2\textbf{r}\left[ \frac{r_s^2}{2}F_n^2(d\chi_{\rho})^2\right]\partial_n^2 f_{\textbf{r}_0}(n(\textbf{r}))
\end{split}
\end{eqnarray}
where in the second line we have used the fact that $\textbf{r}\rightarrow \textbf{r}+d\textbf{r}$ is equivalent to $\textbf{r}_0\rightarrow \textbf{r}_0-d\textbf{r}$ due to homogeneity of the system (of course after averaging over disorder). We can calculate the probability measure of the density, noting that the average value of $f$ should not depend on $\textbf{r}_0$ due to homogeneity of the system. To do averages let us define $D[n]\equiv \prod_{\textbf{r}}dn(\textbf{r})$. The change of the average, due to changing the origin is:
\begin{eqnarray}
\begin{split}
\text{d}\left\langle f_{\textbf{r}_0}\right\rangle &\equiv\text{d} \int d^2\textbf{r}\int D[n]P(\left\lbrace n\right\rbrace)f_{\textbf{r}_0}(n(\textbf{r}))\\
&=\int d^2\textbf{r}\int D[n]P(\left\lbrace n\right\rbrace) \text{d}f_{\textbf{r}_0}(n(\textbf{r})) \\
=\int d^2\textbf{r}&\left\langle \left[ -r_sF_n(\text{d}\chi_{\rho}+\frac{1}{2}\text{d}\chi_n)\right] \partial_n f +\frac{r_s^2}{2}F_n^2(\text{d}\chi_{\rho})^2\partial_n^2f \right\rangle
\end{split}
\end{eqnarray}
Noting again that $\left\langle d\chi_{\rho}=0\right\rangle$, and representing the averages as the integrals over probability measures, and doing integration by parts, we reach at:
\begin{eqnarray}
\begin{split}
\text{d}\left\langle f\right\rangle &=\frac{r_s}{2}\int d^2\textbf{r}\int dn(\textbf{r})[ \partial_n[F_nP(\left\lbrace n\right\rbrace)\text{d}\chi_n]\\
&+ \frac{r_s^2}{2}\partial_n^2[f_n^2P(\left\lbrace n\right\rbrace)(\text{d}\chi_{\rho})^2]]f(n(\textbf{r})) 
\end{split}
\end{eqnarray}

Let us return to the Eq~\ref{MainPnEq} and for convenience define again the used functions as follows:
\begin{equation}
\begin{split}
&\frac{\delta P(\left\lbrace n\right\rbrace)}{\delta n(\textbf{r})}=-\left( \zeta\frac{G_n(\textbf{r})}{F_n(\textbf{r})}+2\partial_n\ln F_n(\textbf{r})\right)P(\left\lbrace n\right\rbrace)\\
&G_n(\textbf{r})=\int d^2\textbf{r}'\frac{n(\textbf{r}')}{|\textbf{r}-\textbf{r}'|^2}\\
& F_n(\textbf{r})=B\frac{\text{sgn}(n(\textbf{r}))\sqrt{|n(\textbf{r})|}}{1-\Omega\left(\text{sgn}(n(\textbf{r}))+\frac{1}{2}\ln (A|n(\textbf{r})|)\right)}
\end{split}
\label{Master}
\end{equation}
in which $A\equiv \frac{4\pi}{(4k_c)^2}$, $B\equiv \frac{2}{\sqrt{\pi}}$ and $\Omega\equiv r_s\beta$. Now let us try the solution:
\begin{equation}
P(\left\lbrace n\right\rbrace)=\exp\left[ -\frac{1}{d^2}\int d^2\textbf{r}'d^2\textbf{r}'' \frac{n(\textbf{r}')n(\textbf{r}'')}{|\textbf{r}'-\textbf{r}''|^2}H(\textbf{r}',\textbf{r}'')\right]
\label{masterH}
\end{equation}
in which $H(\textbf{r}',\textbf{r}'')$ depends on $\textbf{r}'$ and $\textbf{r}''$ via the densities, i.e. $H(\textbf{r}',\textbf{r}'')\equiv H(n(\textbf{r}'),n(\textbf{r}''))$. We obtain:
\begin{equation}
\begin{split}
\frac{\delta P(\left\lbrace n\right\rbrace)}{\delta n(\textbf{r})}&=-\int d^2\textbf{r}'[\frac{2n(\textbf{r}')}{|\textbf{r}'-\textbf{r}|^2}H(\textbf{r}',\textbf{r})\\
&+\frac{n(\textbf{r}')n(\textbf{r})}{|\textbf{r}'-\textbf{r}|^2}\frac{\delta}{\delta n(\textbf{r})}H(\textbf{r},\textbf{r}')]P(\left\lbrace n\right\rbrace)
\end{split}
\end{equation}
in which we have used the symmetry $H(\textbf{r},\textbf{r}')=H(\textbf{r}',\textbf{r})$. The right hand side of the above equation should be equal to the right hand side of Eq.\ref{Master}, i.e.
\begin{equation}
\begin{split}
 -\int d^2\textbf{r}'\left[\frac{\zeta n(\textbf{r}')}{F_n(\textbf{r})|\textbf{r}-\textbf{r}'|^2} + 2\frac{\delta \ln F_n(\textbf{r}')}{\delta n(\textbf{r}')}\delta(\textbf{r}-\textbf{r}')\right]P(\left\lbrace n\right\rbrace)
\end{split}
\end{equation}
Therefore defining $n'\equiv n(\textbf{r}')$, $n\equiv n(\textbf{r})$, $\eta\equiv |\textbf{r}-\textbf{r}'|^2$, $\partial_n\equiv\frac{\delta}{\delta n(\textbf{r})}$, $F\equiv F_n(\textbf{r})$, $F'\equiv F_n(\textbf{r}')$, $s\equiv \text{sgn}(n)$, $s'\equiv \text{sgn}(n')$ and $H\equiv H(\textbf{r},\textbf{r}')$ we obtain:
\begin{equation}
\begin{split}
&2\eta^{-1}n'H+\eta^{-1}nn'\partial_n H=\eta^{-1}\zeta \frac{n'}{F}+2\delta(\textbf{r}-\textbf{r}')\partial_{n'}\ln F'\\
&2\eta^{-1}nH+\eta^{-1}nn'\partial_{n'}H=\eta^{-1}\zeta\frac{n}{F'}+2\delta(\textbf{r}-\textbf{r}')\partial_{n}\ln F
\end{split}
\label{H_eq}
\end{equation}
in which the second equation is obtained by $\textbf{r}\leftrightarrow \textbf{r}'$ symmetry. Let us first consider the case $\textbf{r}\neq\textbf{r}'$ for which we obtain $2H+n\partial_n H=\zeta \frac{1}{F}$ and the same equation for $n\longleftrightarrow n'$. By summing two terms we obtain:
\begin{equation}
H+\hat{L}H=\frac{\zeta}{4}\Sigma
\label{main_H_Eq}
\end{equation}
in which $\hat{L}\equiv\frac{1}{4}(n\partial_n+n'\partial_{n'})$ and $\Sigma\equiv F^{-1}+F'^{-1}$. To continue some relations are essential, namely ($\hat{L}(n^2n'^2)=n^2n'^2$):
\begin{equation}
\begin{split}
&\frac{1}{4}\Sigma=-2\hat{L}\Sigma-\frac{\Omega}{4}\left( \frac{1}{s\sqrt{|n|}}+\frac{1}{s'\sqrt{|n'|}}\right)\\
& \hat{L}(\Sigma)=-\frac{8}{7n^2n'^2}\hat{L}(n^2n'^2\Sigma)-\frac{\Omega}{7}\left( \frac{1}{s\sqrt{|n|}}+\frac{1}{s'\sqrt{|n'|}}\right)
\end{split}
\end{equation}
which is used to re-write Eq. \ref{main_H_Eq} in the following form:
\begin{equation}
\hat{L}(\hat{H})=f(n,n')
\end{equation}
in which $\hat{H}\equiv n^2n'^2\left(H-\frac{2}{7}\zeta\Sigma\right)$ and $f(n,n')\equiv \alpha n^2n'^2\left( \frac{1}{s\sqrt{|n|}}+\frac{1}{s'\sqrt{|n'|}}\right)$ and $\alpha\equiv\frac{\zeta\Omega}{28}$. The general solution of the above equation is as follows:
\begin{equation}
H=\frac{2\zeta}{7}\left[\Sigma + \frac{\Omega}{7}\left(\frac{1}{s\sqrt{|n|}}+\frac{1}{s'\sqrt{|n'|}}\right) \right]+\frac{c}{n^2n'^2}\hat{h}(\frac{n}{n'})
\end{equation}
in which $\hat{h}$ is an arbitrary function having the property $\hat{h}(x)=\hat{h}(\frac{1}{x})$ to restore the $n\leftrightarrow n'$ symmetry of $H$ and $c$ is some constant. It is simply seen that there is no choice but $c=0$. \\
Now let us to turn to the equations~\ref{H_eq}. To satisfy these equations, one should have an asymmetric term to generate the delta function. Satisfying this discontinuity (which is proportional to $\partial_n \ln F(n)$) is a difficult task and the inclusion of $\nabla n$ is needed. Fortunately for the limit $r_s\rightarrow 0$ or $n_i\rightarrow 0$ we can neglect this term ($\zeta\frac{G_n(\textbf{r})}{F_n(\textbf{r})}\gg 2\partial_n\ln F_n(\textbf{r})$) and the term corresponding to the discontinuity becomes very small. Therefore the above analysis is applicable and the final form of $H$ for small densities is as follows:
\begin{equation}
\begin{split}
& H(\textbf{r},\textbf{r}')=\frac{\sqrt{\pi}\zeta}{7}\frac{\text{sgn}(n(\textbf{r}))}{\sqrt{|n(\textbf{r})|}}\times\\
&\left[1+\Omega\left( \frac{2}{7\sqrt{\pi}}-\text{sgn}(n(\textbf{r}))+\ln \frac{4k_c}{\sqrt{4\pi|n(\textbf{r})|}}\right) \right]\\
& + n(\textbf{r})\leftrightarrow n(\textbf{r}')
\end{split}
\end{equation}
With this solution, the exponent of the Eq.~\ref{masterH}, i.e. $\frac{1}{|\textbf{r}-\textbf{r}'|}n(\textbf{r})n(\textbf{r}')H(\textbf{r},\textbf{r}')$ is well-defined and smooth in all range of $n$ and $n'$. 

\section*{Conclusion}\label{conclusion}
In this paper we have investigated the properties of the disorder-averaged dielectric function by analyzing the one- and two- body charge density distribution functions of  mono-layer graphene. For calculating them, we have used the Thomas-Fermi-Dirac (TFD) approach, taking into account the tunable disorder and inter particle interactions on an equal footing. To use the TFD theory for calculating the linear screening, we developed a diagrammatic technique and in the first order approximation in disorder strength, the polarization operator and charge modulation due to external potential have been obtained. The one- and the two- body charge density distribution functions were obtained using some stochastic analysis which carries both the effect of white-noise out-plane disorder and the interaction. The closed form of the functions were obtained for high density and low inter-particle interaction limits. By analyzing the one-body distribution function we found that the charge density fluctuations is proportional to $\zeta^{-1}\sim r_sn_i^2\bar{n}^{-1}$ in which $n_i$ is the disorder strength, $r_s$ is the interaction strength and $\bar{n}$ is the average density of the system, therefore the electron-hole puddles can appear in the low density limit $\bar{n}\rightarrow 0$. By analyzing the two-body charge density distribution function we obtained that:
\begin{equation}
P(n,r;n',0)\propto P_n^{\text{large}\ \zeta}P_{n'}^{\text{large}\ \zeta}\exp\left[-\frac{4\sqrt{2}}{dn_i^2r_s^2}\frac{|n-n'|}{r}\right]
\end{equation} 
in which $ P_n^{\text{large}\ \zeta}$ is the one-body distribution function in the large $\zeta$ limit. Using this function and some other analytical investigations we showed that some power-law behaviors in terms of $\zeta$ raise, specially for $R_{\text{max}}$ which is the length scale at which the static density-density correlation function $G_x(R)$ shows a peak. It was found that $G_x(R)$ falls off logarithmically for large $R$'s, i.e. $G_x(R)^{\text{large}\ R}\propto \log R$. The exponents of the scaling relations have been reported in the text. An interesting effect of disorder and interaction is changing the screened potential both in the Fourier and direct spaces. It was that in addition to $k_F$ a disorder-driven characteristic momentum ($q_{\text{ch}}$) emerges in the system that controls the behaviors of the screened potential. The Fourier component of screened potential changes significantly for small wave numbers. for small enough $\zeta$'s (small densities where EHPs can appear) we have observed an instability in which the imaginary part of the dielectric function becomes negative and some oscillations are observed for the charge density $\delta n(r)$. The oscillations are absent for larger $\zeta$ values (larger densities or smaller inter-particle interactions and disorder strengths). Along with this change, a screening-anti-screening transition occurs in which $\delta n(0)$ changes the sign. The observed instability is certainly affected by the higher order contribution, but we think that it should show itself for small enough densities, since the effect increases in a power-law fashion with $\zeta$.\\
In the last part of the paper we have calculated the total distribution measure of the mono-layer graphene sheets. The resulting equation has been analytically solved for large $\zeta$ limit and was shown to be of the quadratic form. For smaller $\zeta$'s we think that derivatives of charge densities should come to calculations which is the subject of our future research.

\section{Appendix1: Diagrammatic analysis of perturbative disorder in Thomas-Fermi-Dirac theory}\label{appendix1}

\begin{figure}
\centering
  \begin{picture}(300,180)
    \SetArrowScale{2}
    \SetWidth{0.5}
    \Text(60,100)[1]{\huge +}
    \SetColor{Black}
    \ArrowArc[double,sep=2,arrow,arrowinset=0.8,arrowpos=0.4,arrowlength=10](30,60)(15,0,360)
    \ArrowArc[double,sep=2,arrow,arrowinset=0.8,arrowpos=0.4,arrowlength=10](30,140)(15,0,360)
    \ArrowArc[double,sep=2,arrow,arrowinset=0.8,arrowpos=0.4,arrowlength=10](90,60)(15,0,360)
    \Photon[double](30,75)(30,125){5}{6}
    \DashLine(90,75)(90,140){4}
    \GCirc(90,140){10}{0.8}
    \Text(90,140)[1]{\huge $\times$}
    \Text(115,100)[1]{\huge +}
    \Text(40,77)[1]{$x$}
    \Text(40,123)[1]{$x'$}
    \Text(97,77)[1]{$x$}
    \Text(130,87)[1]{$x$}
    \Text(230,87)[1]{$x+\epsilon$}
    \ArrowArc[double,sep=2,arrow,arrowinset=0.4,arrowpos=0.5,arrowlength=13](180,50)(70,45,135)
    \ArrowArc[double,sep=2,arrow,arrowinset=0.4,arrowpos=0.5,arrowlength=13](180,150)(70,225,315)
    \PhotonArc[double](180,100)(50,0,180){5}{6}
    \CCirc(130,100){4}{Black}{Black}
    \CCirc(230,100){4}{Black}{Black}
    \Line(157,82)(157,118)
    \Line(157,82)(198,117)
    \Line(157,87)(192,117)
    \Line(157,93)(189,120)
    \Line(157,100)(182,120)
    \Line(157,107)(174,120)
    \Line(198,83)(198,117)
    \Line(164,82)(198,110)
    \Line(170,82)(198,105)
    \Line(177,82)(198,99)
    \Line(184,82)(198,92)
  \end{picture}
\caption{The digrammatic representation of Eq \ref{average_potential_apendix}.}
\label{fig:averaging1}
\end{figure}
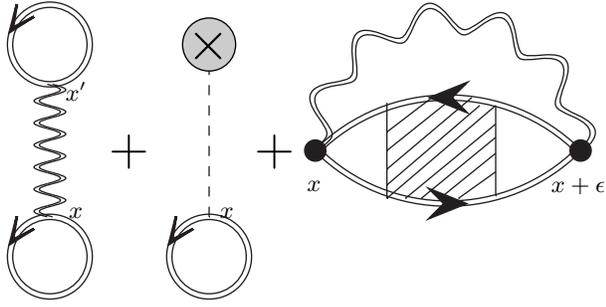

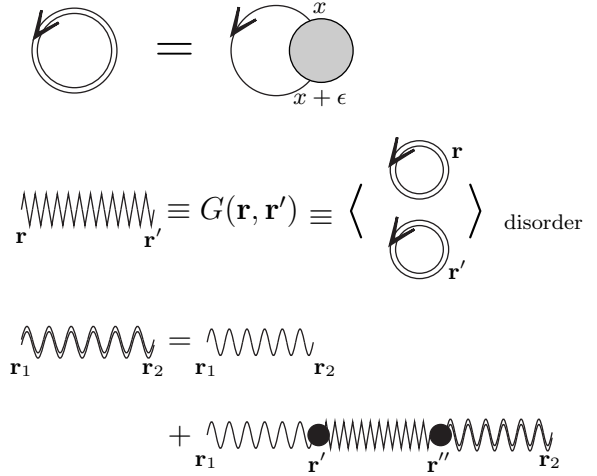
\begin{figure}
\centering
  \begin{picture}(300,220)
    \SetArrowScale{2}
    \SetWidth{0.5}
    \Text(68,190)[1]{\huge =}
    \ArrowArc[double,sep=2,arrow,arrowinset=0.8,arrowpos=0.4,arrowlength=10](30,190)(15,0,360)
    \ArrowArc[arrowpos=0.4,arrowinset=0.8,arrowlength=10](106,190)(17,0,360)
    \GCirc(123,190){12}{0.8}
    \Text(123,207)[1]{$x$}
    \Text(123,173)[1]{$x+\epsilon$}
    \ZigZag(10,130)(60,130){6}{12}
    \Text(10,120)[1]{$\textbf{r}$}
    \Text(60,120)[1]{$\textbf{r}'$}
    \Text(90,130)[1]{\large $\equiv G(\textbf{r},\textbf{r}')$}
    \Text(130,130)[1]{\large $\equiv$ \huge $ \left\langle \right. $}
    \ArrowArc[double,sep=2,arrow,arrowinset=0.8,arrowpos=0.4,arrowlength=10](160,145)(10,0,360)
    \ArrowArc[double,sep=2,arrow,arrowinset=0.8,arrowpos=0.4,arrowlength=10](160,115)(10,0,360)
    \Text(200,130)[1]{\huge $\left. \right\rangle$ \large $_{\text{disorder}} $}
    \Text(175,152)[1]{$\textbf{r}$}
    \Text(175,108)[1]{$\textbf{r}'$}
    \Photon[double](10,80)(60,80){5}{6}
    \Text(70,80)[1]{\large $=$}
    \Photon(80,80)(120,80){5}{6}
    \Text(70,45)[1]{\large $+$}
    \Photon(80,45)(120,45){5}{6}
    \ZigZag(124,45)(164,45){5}{12}
    \Photon[double](170,45)(210,45){5}{6}
    \CCirc(122,45){4}{Black}{Black}
    \CCirc(168,45){4}{Black}{Black}
    \Text(122,35)[1]{$\textbf{r}'$}
    \Text(168,35)[1]{$\textbf{r}''$}
    \Text(10,70)[1]{$\textbf{r}_1$}
    \Text(60,70)[1]{$\textbf{r}_2$}
    \Text(80,70)[1]{$\textbf{r}_1$}
    \Text(125,70)[1]{$\textbf{r}_2$}
    \Text(80,35)[1]{$\textbf{r}_1$}
    \Text(210,35)[1]{$\textbf{r}_2$}
  \end{picture}
\caption{The diagrammatic expansion of the full density $n(\textbf{r})$ (the first line). The gray circle shows the full electronic propagator which has become localized due to coarse graining. The diagrammatic representation of the mediator $G(\textbf{r},\textbf{r}')$ (second line). The diagrammatic expansion of the full interaction line corresponding to the mediator $G(\textbf{r},\textbf{r}')$ (the last line).}
\label{fig:averaging2}
\end{figure}

In this appendix we analyze the disorder averaging and its diagrammatic interpretation. A complexity of the graphene system arises from the fact that the potentials are long-range for this system and the short-range techniques for treating the impurities \cite{sadovskii2006diagrammatics} are not applicable. $G(\textbf{r})$ has a non-trivial effect on the energy expectation value via \cite{fetter2012quantum}:
\begin{equation}
\begin{split}
\left\langle \left\langle \Omega\right| \hat{V}\left| \Omega\right\rangle \right\rangle & = \frac{1}{2}\int d^2\textbf{r}d^2\textbf{r}' V(\textbf{r}-\textbf{r}')\left[ i\hbar\Pi(\textbf{r},t;\textbf{r}',t)_{\text{connected}}\right. \\
& \left. +G(\textbf{r},\textbf{r}')-\delta\left(\textbf{r}-\textbf{r}'\right)n(\textbf{r})\right]
\end{split}
\label{average_potential_apendix}
\end{equation}
in which $\Pi(\textbf{r},t;\textbf{r}',t)_{\text{connected}}$ is the second term in Fig. \ref{fig:averaging} which also contains the disorder interaction lines \cite{sadovskii2006diagrammatics}. The effect of the second term ($G(\textbf{r})$) on the ground state energy is analyzed in this appendix in the Thomas-Fermi-Dirac (TFD) theory. This theory is a coarse-grained approximation for an electronic system, so that all the contributions are localized in a region in the close vicinity of the original spatial point, except the Hartree term, as well as the disorder term which are non-local terms. The Hartree term arises from the second term in the right hand side of Eq. \ref{average_potential_apendix}. These have been shown in the Fig. \ref{fig:averaging1} in which the first term is the Hartree term, the second one is the external disorder interaction and the last term shows the other terms which are local in the coarse-grained system (to show this, we used $x+\epsilon$ to mention the close vicinity of $x$). The double (coulomb and electronic) lines show the full propagators. The double-line circles show the full electronic density which are the solution of the TFD equations and have been shown in Fig. \ref{fig:averaging2}. Full lines have simultaneously the disorder and coulomb interaction lines. 

Now let us consider the screening of the potential in one-loop level. With an appropriate contractions in the first order perturbation, we have in the real space:
\begin{equation}
\begin{split}
& v(\textbf{r}_1-\textbf{r}_2)=v^0(\textbf{r}_1-\textbf{r}_2)+\\
& \frac{i}{\hbar}\int\int d^2\textbf{r}'d^2\textbf{r}''v^0_{\textbf{r}_1,\textbf{r}'}v^0_{\textbf{r}'',\textbf{r}_2}G(\textbf{r}'-\textbf{r}'')+\text{higher order terms}.
\end{split} 
\end{equation} 
in which $v^0_{\textbf{r}_1,\textbf{r}_2}\equiv v^0(\textbf{r}_1-\textbf{r}_2)$. In the coarse-grained version, this term (and also higher order terms containing $G(\textbf{r},\textbf{r}')$) survives and plays an important role. In the third line of the Fig. \ref{fig:averaging2} we have shown the infinite series of the above equation, in which the most left bare coulomb line is replaced by the full line. By taking Fourier transform of this series, and defining $G(q)=\int d^2\textbf{r}e^{iq.\textbf{r}}G(\textbf{r})$, we reach to the following equation:
\begin{equation}
v(q)=\frac{v_q^0}{\epsilon(q,\omega)}
\end{equation}
in which the dielectric function is determined to be $\epsilon(q,\omega)=1+\frac{i}{\hbar}v_q^0G(q)\delta(\omega)$. In the real systems, the time scales up to the time required for a photon to pass through the sample can be interpreted a instantaneous, i.e. $\delta(\omega)\sim \tau_{\text{ch}}\equiv \frac{d}{c}$. Using this fact and the relation \ref{G_x} and the fact that $\epsilon(G_q=0)=\kappa_S$, one can show that:
\begin{equation}
\begin{split}
\epsilon(q,\omega) & \approx \kappa_S+i\frac{\gamma_0\kappa_S}{dr_s^3}\left( \frac{v_F}{c}\right) \frac{\tilde{G}(\tilde{q})}{q}\\
& = \kappa_S\left( 1+i\frac{\tilde{q}_0}{\tilde{q}}\tilde{G}(\tilde{q})\right) 
\end{split}
\end{equation}
in which $\tilde{q_0}\equiv \frac{\gamma}{n_id^2r_s^5}\left(\frac{v_F}{c}\right)$, $\gamma_0\equiv 128$ and $\gamma\equiv 512\sqrt{2/\pi}$ and $\tilde{q}\equiv \frac{n_i\zeta'}{\alpha}q$ as defined in the SEC. \ref{TwoBody}. For a system with $d^2n_i\sim 1$ and $r_s=0.8$, $\tilde{q}_0\sim 23.5$. One may concern about the limits $n_i\rightarrow 0$ or $r_s\rightarrow 0$ for which $\tilde{q}_0\rightarrow \infty$. This actually is not a problem since for these quantities $\tilde{q}\rightarrow \infty$ for which $\tilde{G}$ is zero. In fact in this limits the imaginary part of $\epsilon(q,\omega)$ vanishes (as is seen in the first line of the above equation) as expected. Therefore the normalized screened potential, defined by $V_q\equiv \frac{\alpha}{4\pi e^2n_i\zeta'}v_q=\frac{dr_s^2n_i}{16\sqrt{2\pi}e^2}v_q$ (so that $V_q^0=\tilde{q}^{-1}$) takes the following form:
\begin{equation}
\kappa_S\text{Re}[V_q]=\frac{V_q^0}{1+\frac{\tilde{q}_0^2}{\tilde{q}^2}\tilde{G}(\tilde{q})^2}
\end{equation}
This result should be compared with the relation $\epsilon(q)=\epsilon_0- \frac{qe^2}{2\omega_p(q)}\ln\left(\frac{2|\mu|-\omega_p(q)}{2|\mu|+\omega_p(q)}\right)$ which was obtained by Shung that yields the change of the dielectric function and leads to the relation~\cite{Shung1986}:
\begin{equation}
\frac{v_0(q)}{\epsilon(q,0)}=\frac{1}{\epsilon_0}\frac{2\pi e^2}{q+q_{\text{TF}}}
\end{equation}
in which $q_{\text{TF}}\equiv 4\pi e^2k_F/v_F\epsilon_0$. In our case one can easily show that:
\begin{equation}
\begin{split}
\frac{v_{\text{sc}}(r)}{4\pi e^2} & =\frac{1}{4\pi e^2}\int d^2q \frac{v_0(q)}{\epsilon(q,\omega)}e^{iq.r}\\
&=\frac{\sqrt{\pi}dn_i^2r_s^2}{4\sqrt{2}\kappa_S}f_1(R)\\
\frac{\delta n(r)}{4\pi e^2} & = \frac{1}{4\pi e^2}\int d^2q \frac{v_0(q)G(q)}{\epsilon(q,\omega)}e^{iq.r}\\
& =\frac{8\sqrt{\pi}}{\sqrt{2}\kappa_Sd}\left( \frac{n_i}{r_s}\right)^2f_2(R)
\end{split}
\end{equation}
in which $v_{\text{sc}}(r)$ is the screened potential which is obtained from $\epsilon(q,\omega)$ and:
\begin{equation}
\begin{split}
f_1(R)&\equiv 2\pi\int_0^{\infty}\frac{J_0(\tilde{q}R)}{1+\left( \frac{\tilde{q}_0}{\tilde{q}}\tilde{G}(\tilde{q})\right)^2}d\tilde{q}\\
f_2(R)&\equiv 2\pi\int_0^{\infty}\frac{\tilde{G}(\tilde{q})J_0(\tilde{q}R)}{1+\left( \frac{\tilde{q}_0}{\tilde{q}}\tilde{G}(\tilde{q})\right)^2}d\tilde{q}
\end{split}
\end{equation}
It is notable that in the limit $n_i\rightarrow 0$ one obtains the trivial result $v_{\text{sc}}(r)\rightarrow v_0(r)$.

\section{Appendix2}\label{appendix2}

In this short appendix we concentrate on calculating $\text{d}\chi_n(0)\equiv \nabla \chi_n(0).\text{d}\textbf{r}$ and $(\text{d}\chi_{\rho}(d))^2\equiv (\nabla \chi_{\rho}(d).\text{d}\textbf{r})^2$. Writing $\text{d}\chi_{\rho}(d)$ as $\int d^2\textbf{r}^{\prime}\rho(\textbf{r}^{\prime})\left[\left(|\textbf{r}+\text{d}\textbf{r}-\textbf{r}^{\prime}|+d^2 \right)^{-1/2} -\left(|\textbf{r}-\textbf{r}^{\prime}|+d^2\right)^{-1/2}\right]$ we have:
\begin{eqnarray}
\begin{split}
&\left\langle (\text{d}\chi_{\rho})^2\right\rangle =\int d^2\textbf{r}'d^2\textbf{r}''\\
&\left[ (|\textbf{r}-\textbf{r}^{\prime}|^2+d^2)(|\textbf{r}-\textbf{r}''|^2+d^2)\right]^{-\frac{1}{2}}\times\\
&\left\langle \left( \rho(\textbf{r}^{\prime}+\text{d}\textbf{r})-\rho(\textbf{r}^{\prime})\right) \left(\rho(\textbf{r}''+d\textbf{r})-\rho(\textbf{r}'')\right) \right\rangle
\end{split}
\end{eqnarray}
By expanding the integrand in terms of $\text{d}\textbf{r}$, i.e. $\left(|\textbf{r}-\textbf{r}^{\prime}-\text{d}\textbf{r}|^2+d^2\right)^{-1/2}\simeq \left(|\textbf{r}-\textbf{r}^{\prime}|^2+d^2\right)^{-1/2}+\text{d}\textbf{r}.\nabla \left(|\textbf{r}-\textbf{r}^{\prime}|^2+d^2\right)^{-1/2}$, we obtain
\begin{eqnarray}
\begin{split}
&\left\langle (\text{d}\chi_{\rho})^2\right\rangle =2(n_id)^2\int d^2\textbf{r}^{\prime}\frac{(\textbf{r}-\textbf{r}^{\prime}).\text{d}\textbf{r}}{(|\textbf{r}-\textbf{r}^{\prime}|^2+d^2)^2}
\end{split}
\end{eqnarray}
By replacing $\cos^2\theta\equiv\frac{1}{2}$, we see that $\left\langle (\text{d}\chi_{\rho})^2\right\rangle =\frac{\pi dn_i^2}{2\sqrt{2}}\text{d}r$. Note that $\text{d}\chi_n(0)\simeq \frac{1}{\sqrt{2}}G_n \text{d}r$ in which $G_n=\int d^2\textbf{r}^{\prime}\frac{n(\textbf{r}^{\prime})}{|\textbf{r}-\textbf{r}^{\prime}|^2}$.

\section{Appendix3}\label{appendix3}

This appendix has been devoted to the calculation of functional derivative of $P$. Our functional derivatives are a bit different from the common definitions which needs clarification. Let us consider the integral 
\begin{equation}
\text{d}\left\langle f_{\textbf{r}_0}\right\rangle=\int d^2\textbf{r}\int D[n]\text{d}X(n(r))P\left( \left\lbrace n\right\rbrace \right) \partial_nf_{\textbf{r}_0}(n(r))
\end{equation}
To define the derivatives we mesh the system by squares of sizes $d\times d$, so that the above integral becomes:
\begin{equation}
\begin{split}
\text{d}\left\langle f_{\textbf{r}_0}\right\rangle=&\sum_id^2\int dn_0dn_1dn_2...dn_i... P(n_0,n_1,n_2,...,n_i,...)\times\\
& \text{d}X(n_i)\partial_{n_i}f(n_i,n_0)\\
=&-d^2\sum_i\int dn_0dn_1dn_2...dn_i...  \\
& \partial_{n_i}\left[ P(n_0,n_1,n_2,...,n_i,...)\text{d}X(n_i)\right] f(n_i,n_0)
\end{split}
\end{equation}
The second equality was obtained using integration by parts. If we perform the integration of the above equation over $D[n]'\equiv dn_1...dn_{i-1}dn_{i+1}...$, the equation governing $f(\textbf{r}_0,\textbf{r})=\left\langle n(\textbf{r}_0,n(\textbf{r}))\right\rangle$ presented in SEC.~\ref{TwoBody} is obtained, noting that:
\begin{equation}
\begin{split}
P(n,\textbf{r};n',\textbf{r}_0)&=\int D[n]' P\left( \left\lbrace n\right\rbrace \right)\\
&=\int \prod_{j\ne 0,i}dn_j P(n_0,n_1,...,n_i,...)
\end{split}
\end{equation}
in which $n_j\equiv n_j(\textbf{r}_j)$ and  $\textbf{r}\equiv\textbf{r}_i$. On the other hand, as explained in the text, we consider $P\left( \left\lbrace n\right\rbrace \right) $ to be of the form $\exp\left[-\frac{1}{d^2}\int d^2\textbf{r}'d^2\textbf{r}''n(\textbf{r}')n(\textbf{r}'')H(\textbf{r}',\textbf{r}'')/|\textbf{r}'-\textbf{r}''| \right]$. Therefore we have ($H_{i,j}\equiv H(n(\textbf{r}_i),n(\textbf{r}_j))$, $r_{i,j}\equiv|\textbf{r}_i-\textbf{r}_j|$ and noting that $\int\int d^\textbf{r}d^2\textbf{r}'\rightarrow d^4\sum{i,j}$):
\begin{equation}
\begin{split}
P & \sim \exp\left[-\frac{1}{d^2}\sum_{i,j}d^4\frac{n_in_j}{r_{i,j}}H_{i,j}\right]\\
&=\prod_{i\neq j}\exp\left[- d^2\frac{n_in_j}{r_{i,j}}H_{i,j}\right]
\end{split}
\end{equation}
which yields 
\begin{equation}
\begin{split}
\partial_{n_i} \ln\left( P\left( \left\lbrace n\right\rbrace \right)\right) &= -d^2\sum_{j}\left[ \frac{n_j}{r_{i,j}}\left( 2+n_i\partial_{n_i}\right) H_{i,j}\right]\\
\text{in continuum limit} & \longrightarrow -\int d^2\textbf{r}' \frac{n(\textbf{r}')}{|\textbf{r}-\textbf{r}'|}(2+n(\textbf{r}')\partial_{n(\textbf{r})})\\
&\times H(n(\textbf{r}),n(\textbf{r}'))
\end{split}
\end{equation}
which coincides the calculations in SEC~\ref{TotalDistribution}. 

\bibliography{refs}

\end{document}